\documentclass[aps,twocolumn,showkeywords,floatfix,superscriptaddress]{revtex4}
\usepackage{graphicx}
\usepackage{hyperref}
\usepackage{amsfonts}
\usepackage{amsmath,bm,amssymb}
\usepackage{mathtools}
\usepackage{comment}
\usepackage{xr-hyper}
\usepackage{hyperref}
\usepackage{placeins}

\begin{document}

\title{Proximity effect in a superconductor-topological insulator heterostructure based on first principles}

\author{Kyungwha Park}
\email{Corresponding author: kyungwha@vt.edu}
\affiliation{Department of Physics, Virginia Tech, Blacksburg, Virginia 24061, USA}
\author{Gabor Csire}
\email{gabor.csire@icn2.cat}
\affiliation{Catalan Institute of Nanoscience and Nanotechnology (ICN2),
CSIC, BIST, Campus UAB, Bellaterra, Barcelona 08193, Spain}
\author{Balazs Ujfalussy}
\email{ujfalussy.balazs@wigner.mta.hu}
\affiliation{Wigner Research Centre for Physics, Budapest, P.O. Box 49 H-1525, Hungary}

\date{\today}

%\setcitestyle{super}

%define onlinecite for non-revtex paper
%\newcommand{\onlinecite}[1]{\hspace{-1 ex} \nocite{#1}\citenum{#1}}
\externaldocument[S-]{SI_PRB}
%
%\section{a-b-s-t-r-a-c-t} KP Apr 9, 2020: 148 words below. We can afford only two more words if needed.
\begin{abstract}
Superconductor-topological insulator (SC-TI) heterostructures were proposed to be a possible platform to realize and control Majorana zero-modes. Despite experimental signatures indicating their existence, univocal interpretation of the observed features demands theories including realistic electronic structures. To achieve this, we solve the Kohn-Sham-Dirac-Bogoliubov-de Gennes equations for ultrathin Bi$_2$Se$_3$ films on superconductor
PdTe, within the fully relativistic Korringa-Kohn-Rostoker method, and investigate quasiparticle spectra as a function of chemical potential and film thickness. We find a strongly momentum-dependent proximity-induced gap feature where the gap sizes highly depend on characteristics of the TI states. The interface TI Dirac state is relevant to the induced gap only when the chemical potential is close to the Dirac-point energy. Otherwise, at a given chemical potential, the largest induced gap arises from the highest-energy quantum-well states, whereas the smallest gap arises from the TI topological surface state with its gap size depending on the TI pairing potential.
\end{abstract}

%\section{m-a-i-n}
%\pacs{TBA}

\maketitle

\section{Introduction}

Superconducting proximity effects in nanoscale heterostructures have been studied in various experimental platforms since proposals \cite{Kitaev2001,Fu2008,Alicea2012} on realization of Majorana zero-modes in them. Representative systems are superconductor(SC)-topological insulator(TI) heterostructures \cite{FYang2012,MXWang2012,SYXu2014,JPXu2014,QLHe2014,JPXu2015,Flototto2018,HHSun2017}
and semiconductor nanowires or ferromagnetic chains on SC substrates
\cite{Mourik2012,Gul2017,HZhang2018,NadjPerge2014,Feldman2017,Ruby2017,Morales2019}.
Majorana zero modes remain robust over local environmental perturbations due to the topological nature, which allows their applications
to quantum computation \cite{Kitaev2003}. Experimental data has exhibited signatures of Majorana zero modes, yet
there remains debate on whether other possibilities can be completely excluded for them. This doubt is raised from
inconsistent experimental findings and current theoretical limitations.

% Experimental side inconsistency here

In SC-TI heterostructures, it was predicted that the topological TI interface state creates proximity-induced topological superconductivity
where the edge state or the vortex lattice can host Majorana zero-modes. Keeping this in mind, experiments
were performed on Bi$_2$Se$_3$(111) or Bi$_2$Te$_3$(111) films ($<$ 10 nm) grown on NbSe$_2$ or Nb substrates, using scanning tunneling miscroscopy/spectroscopy (STM/S) \cite{MXWang2012,JPXu2014,JPXu2015} and angle-resolved photoemission spectra (ARPES) \cite{SYXu2014,Flototto2018}.
A proximity-induced gap was measured when the Fermi level lies in the TI conduction band. The proximity-induced gap was shown to decrease
with increasing TI film thickness. However, both characteristics of the gap and its dependence on film thickness remain elusive
and inconsistent among different groups. The induced surface gap was reported to reach 15-40\% of the bulk SC gap
\cite{MXWang2012,JPXu2014,Flototto2018} even for thick TI films where the interface and top-surface Dirac states do not hybridize.
Intriguingly, transport experiments on Bi$_2$Se$_3$ with a Pb overlayer showed zero resistivity and a proximity-induced gap about 1 $\mu$m
away from the interface \cite{FYang2012}.

% theoretical limitations

Theoretical efforts have been made to understand the proximity effect in SC-TI heterostructures, using the Bogoliubov-de Gennes Hamiltonian
for SC and effective models for TI \cite{Fu2008,Alicea2012,Stanescu2010,Lababidi2011,Schaffer2013,CKChiu2016}.
The TI model Hamiltonian ranges from the Fu-Kane-Mele model \cite{Fu2007} to the
$k \cdot p$ model \cite{HZhang2009} with parameter values from calculated band structures using density-functional theory (DFT).
For chemical potential away from the Dirac point, triplet pairing was predicted \cite{Fu2008,Stanescu2010}, and proximity-induced pairing
types were classified based on symmetries \cite{Schaffer2013}. Despite this success, the current theoretical approaches have many limitations
due to the lack of information on realistic band structures and interface effects as well as insufficient treatment of quasi-two-dimensional
experimental systems. The reported theoretical studies either neglected quantum-well states (QWS) or included generic QWS, although the
experimental Fermi level typically lies in the TI conduction band, and they oversimplified the Fermi surfaces despite their importance in
the induced gap features.

% Summarize our key result here.

We investigate quasiparticle spectra of heterostructures comprised of thin Bi$_2$Se$_3$ overlayers on a $s$-wave SC PdTe substrate, by solving
the Dirac-Bogoliubov-de Gennes (DBdG) equations \cite{Capelle1999} using the fully relativistic screened Korringa-Kohn-Rostoker (SKKR) Green's
function method \cite{Csire2015,Csire2018} within DFT. We calculate the band structure of Andreev bound states and determine proximity-induced
gap features in the TI layers, considering two different TI pairing potentials, as the overlayer thickness and chemical potential are varied.
Several different types of induced gaps appear depending on the characteristics of the TI states for a given overlayer thickness and chemical potential. The induced gap arising from the interface TI Dirac state is the largest, while the gap from the top-surface TI state is smallest,
independently of the thickness, chemical potential, and TI pairing potential. The induced gap associated with a given QWS type increases as the chemical potential decreases. The induced gap sizes do not have a spatial dependence except for the top-surface TI Dirac state for thick TI
films. The induced gap size from the top-surface TI Dirac state is highly susceptible to the TI pairing potential, although that is not the case
for the induced gap from the interface TI Dirac state.

We first discuss our systems of interest in Sec. II and present our first-principles calculations of the electronic structure of
the heterostructures in normal state as a function of overlayer thickness in Sec.III.A. We then show our calculated spectral functions of the heterostructures in SC state with two different TI pairing potentials as a function of overlayer thickness and chemical potential in Sec.III.B.
We make brief comparison of our results with relevant experimental data in Sec.III.C and make a conclusion in Sec.IV.

%%%%%%%%%%%%%%%%%%%%%%%%%%%%%%%%%%%%%%%%%%%%%%%%%%%%%%%%%%%%%%%%%%%%%%%
\section{Systems of interest}\label{sec2:systems}

\begin{figure*}[h!]
\begin{center}
\includegraphics[width=0.7\textwidth]{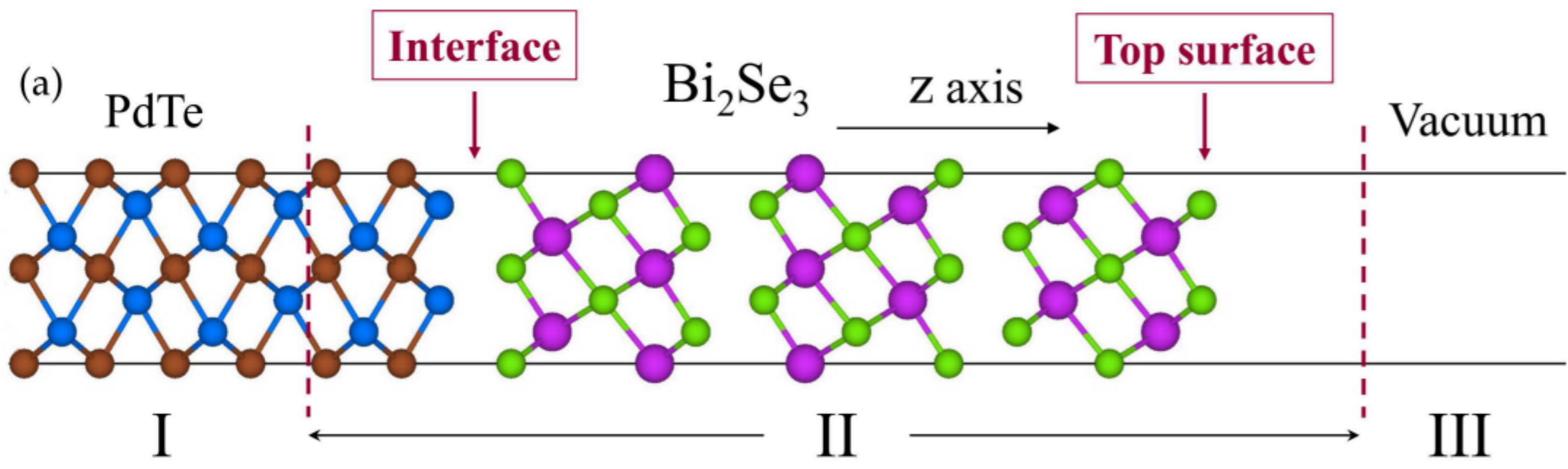}

\vspace{0.5truecm}

\includegraphics[width=0.3\textwidth]{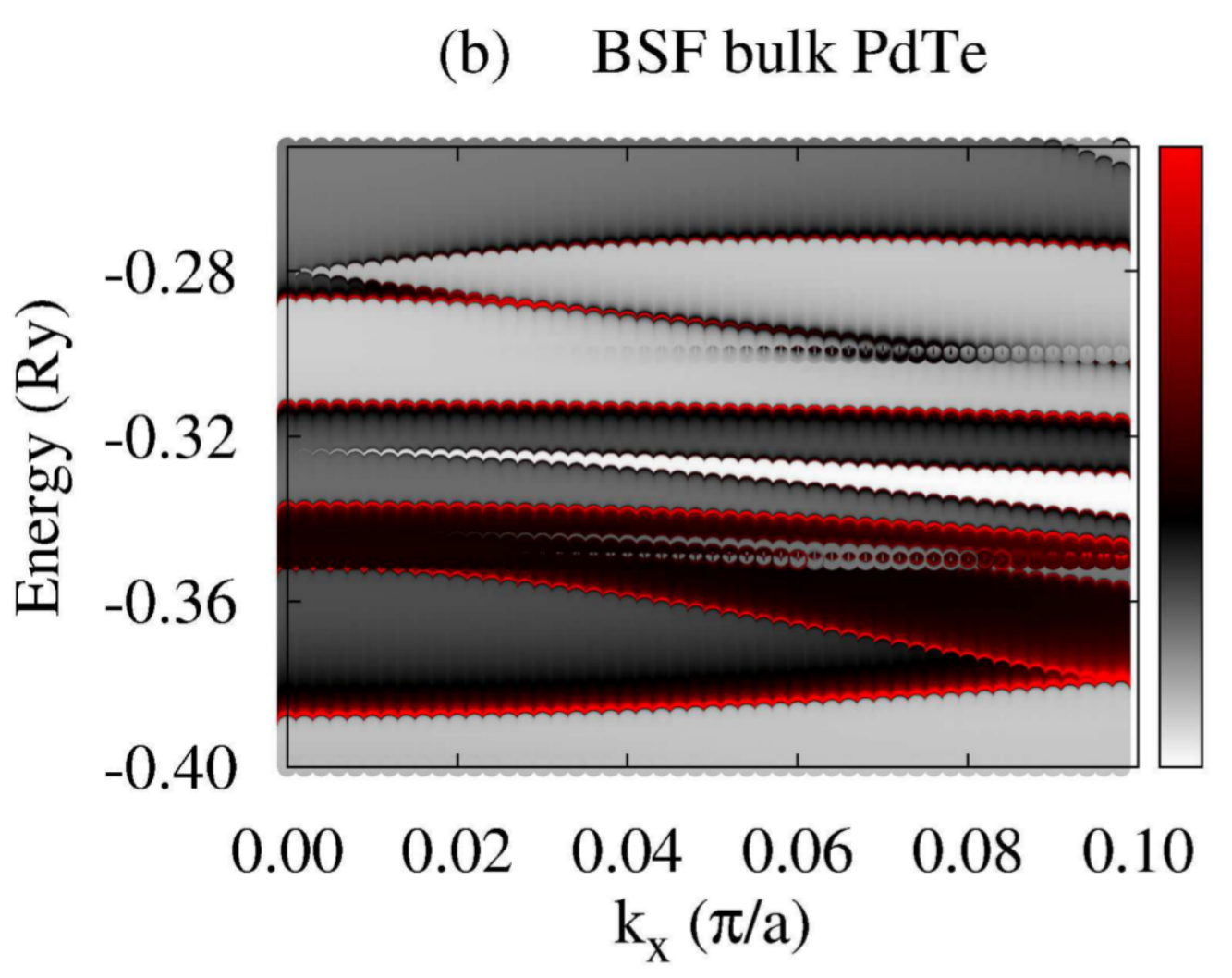}
\hspace{0.2truecm}
\includegraphics[width=0.28\textwidth]{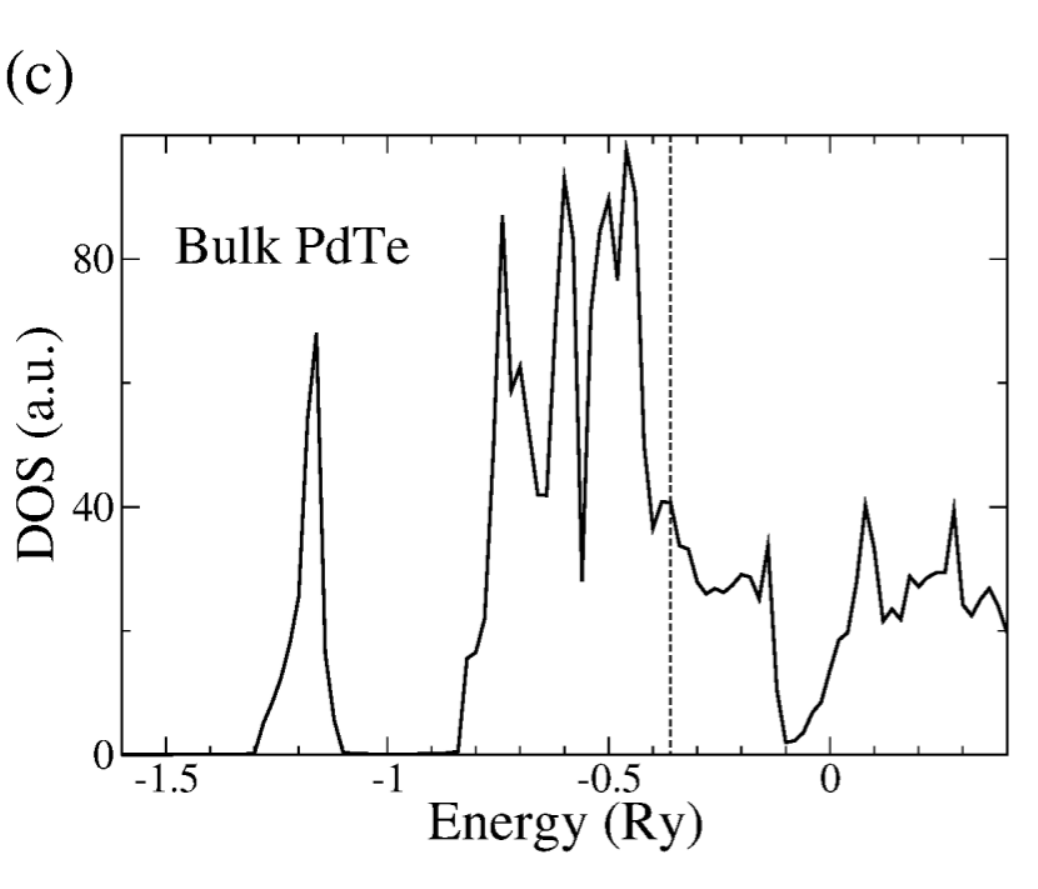}
\hspace{0.2truecm}
\includegraphics[width=0.3\textwidth]{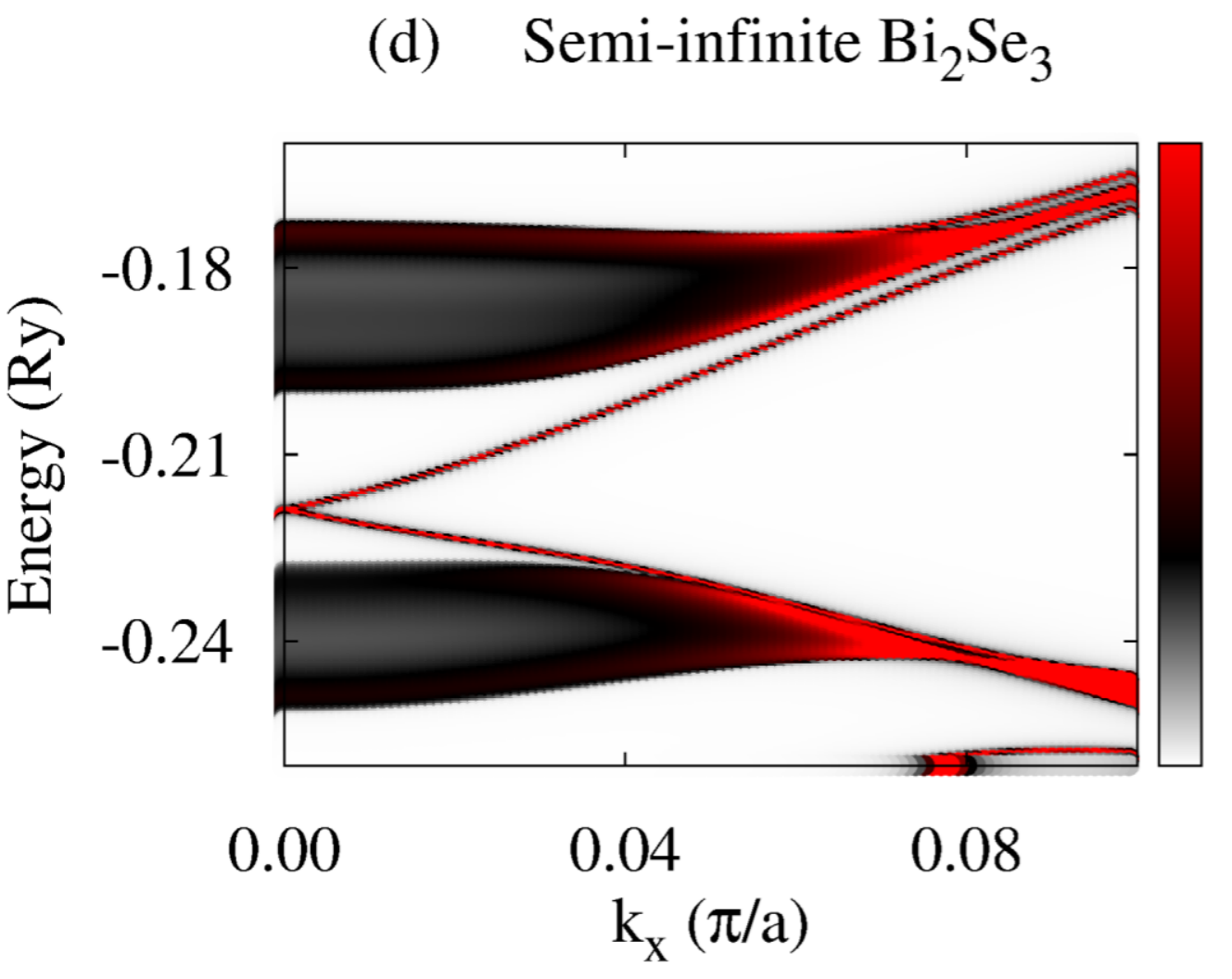}
\caption[Bi2Se3]{(a) Schematic diagram of the SC-TI heterostructures for our SKKR-based simulations.
A Bi$_2$Se$_3$(111) film is overlaid on PdTe(001). Region II (interface region) consists of Bi$_2$Se$_3$ films
of 1-6 QLs (about 1-6 nm) and 4 PdTe atomic layers, while region I and III are semi-infinite PdTe and vacuum layers, respectively.
Color code is as follows: Pd (brown), Te (blue), Se (green), Bi (purple).
(b) SKKR-calculated BSF contours and (c) DOS for bulk PdTe (dashed vertical line: $E_F$), and (d) SKKR-calculated BSF contours for
a semi-infinite Bi$_2$Se$_3$ system. In (b) and (d) the BSF weight increases from white to red in vertical color bars. Dirac surface states
are shown within the bulk band gap in (d). In the SC-TI heterostructures (Fig.~\ref{fig:Normal_BSF}) the TI Dirac point is found near
$-0.29$~Ry due to a shift of the Madelung potential.}
\label{fig:geo}
\end{center}
\end{figure*}

To incorporate realistic electronic structures in the superconducting proximity effect, we simulate Bi$_2$Se$_3$ films of 1-6 quintuple
layers (QLs) overlaid on SC PdTe [Fig.~\ref{fig:geo}(a)] within the fully relativistic SKKR method \cite{Csire2015,Csire2018}. We consider 
such heterostructures due to
the following advantages: (i) good lattice match at the interface; (ii) reasonable SC transition temperature of PdTe; (iii) a single Dirac cone
of Bi$_2$Se$_3$ at a given surface \cite{HZhang2009}; (iv) experimental data on SC-Bi$_2$Se$_3$ heterostructures (despite different SC substrates) \cite{FYang2012,MXWang2012,SYXu2014,JPXu2014,Flototto2018}. Bulk Bi$_2$Se$_3$ has a rhombohedral structure (space group 166, $R\bar{3}m$) with experimental lattice constants $a=4.143$ and $c=28.636$~\AA~\cite{Nakajima1963}. As shown in Fig.~\ref{fig:geo}, Se-Bi-Se-Bi-Se atomic layers form
one QL (about 1 nm) along the [111] direction, and individual QLs are bonded via weak van der Waals interaction.
The (111) surface has a hexagonal in-plane lattice with a lattice constant $a=4.143$~\AA.~
Bulk PdTe has a NiAs-type hexagonal structure (space group 194, $P6_3/mmc$) with experimental lattice constants $a=4.152$ and
$c=5.671$~\AA~\cite{Karki2012}. When a Se-terminated Bi$_2$Se$_3$(111) film is interfaced with a Te-terminated PdTe(001) substrate, the in-plane
lattice mismatch between Bi$_2$Se$_3$ and PdTe is about 0.2\%. PdTe is a type-II SC with critical temperature of 4.5~K \cite{Karki2012}
and its SC gap $\Delta_{\rm{PdTe}}$ is about 0.71~meV at zero temperature \cite{Karki2012}. The electron-phonon coupling of PdTe is
1.4 \cite{Karki2012}, whereas the reported electron-phonon coupling of Bi$_2$Se$_3$ has an upper bound of 0.43 \cite{XZhu2012}.

As required in the SKKR formalism, the SC-TI heterostructures are divided into three regions I-III (Fig.~\ref{fig:geo}(a)). Region I
and III are semi-infinite PdTe and vacuum layers, respectively. Region II consists of Pd-Te-Pd-Te atomic layers overlaid with Bi$_2$Se$_3$
layers (1-6 QLs) and 4-6 vacuum layers on top. The heterostructures have two-dimensional translational symmetry with the lattice
constant of Bi$_2$Se$_3$, 4.143~\AA.~Our choice of the in-plane lattice constant is made to avoid an effect of strain on the topological
surface states \cite{Young2011,WJYang2018}. The out-of-plane lattice constant for PdTe is slightly expanded to conserve the volume of PdTe.
Otherwise, experimental lattice constants are used. We consider the band structure only along the $\Gamma-K$ direction, $k_x$ axis, in this work. 
%since induced-gap features for other directions are expected to be similar.

%%%%%%%%%%%%%%%%%%%%%%%%%%%%%%%%%%%%%%%%%%%%%%%%%%%%%%%%%%%%%%%%%%%%%%
\section{Results and Discussion}

In multiple scattering theory the band structure is obtained from site-dependent Bloch spectral functions (BSF)
$A^{\rm B}_i({\cal E},{\bf k}_{\parallel})$, where $i$ denotes the $i$-th atomic site and ${\cal E}_n({\bf k}_{\parallel})$ is the
$n$-th band energy at in-plane momentum ${\bf k}_{\parallel}$. We calculate the BSF for the $i$-th site (located at ${\bf r}_i$) from its
retarded Green's function $G_{i}^{+}({\cal E},{\bf r}_i,{\bf k}_{\parallel})$:
\begin{equation}
A_i^{\rm B}({\cal E},{\bf k}_{\parallel}) =
-\frac{1}{\pi} {\rm{Im}} {\rm{Tr}} \int d{\bf r}_i G_{i}^{+}({\cal E},{\bf r}_i,{\bf k}_{\parallel}).
\end{equation}
Density of states (DOS) within the SKKR method is obtained from an integral of the BSF over ${\bf k}_{\parallel}$,
i.e., ${\cal D}({\cal E})=\sum_i \int d{\bf k}_{\parallel} A_i^{\rm B}({\cal E},{\bf k}_{\parallel})$.
A detailed description of solving the DBdG equations within the fully relativistic SKKR method can be found in Refs.\onlinecite{Csire2015,Csire2018}.
A brief method description with parameter values for our simulations is shown in the Supplemental 
Material (SM).

\subsection{Electronic structure of normal state}\label{sec3:normal}

% KKR results
% comparison with VASP
%\subsection{PdTe and Bi$_2$Se$_3$}

\begin{figure*}[!h]
\begin{center}
\includegraphics[width=0.31\textwidth]{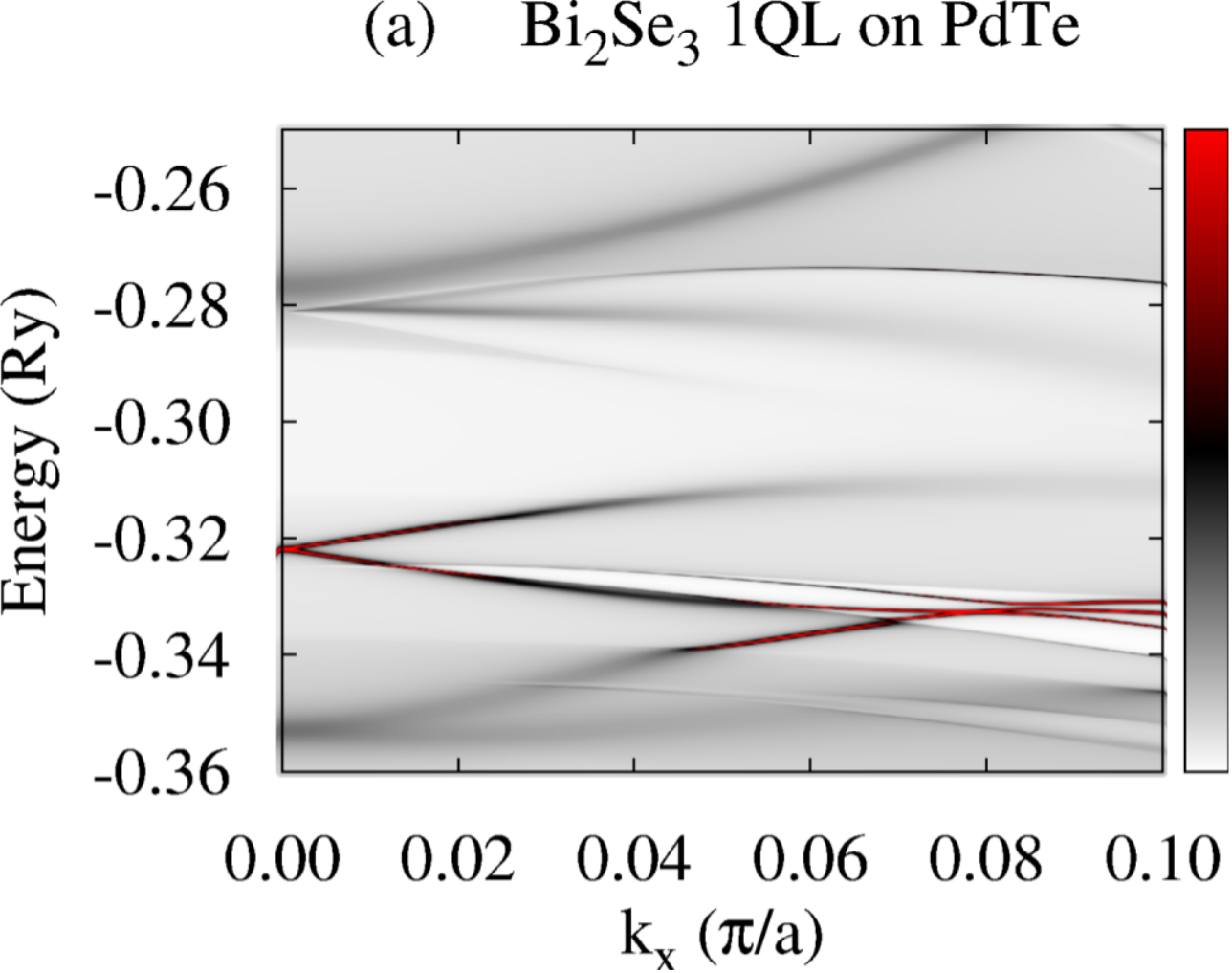}
\hspace{0.2truecm}
\includegraphics[width=0.31\textwidth]{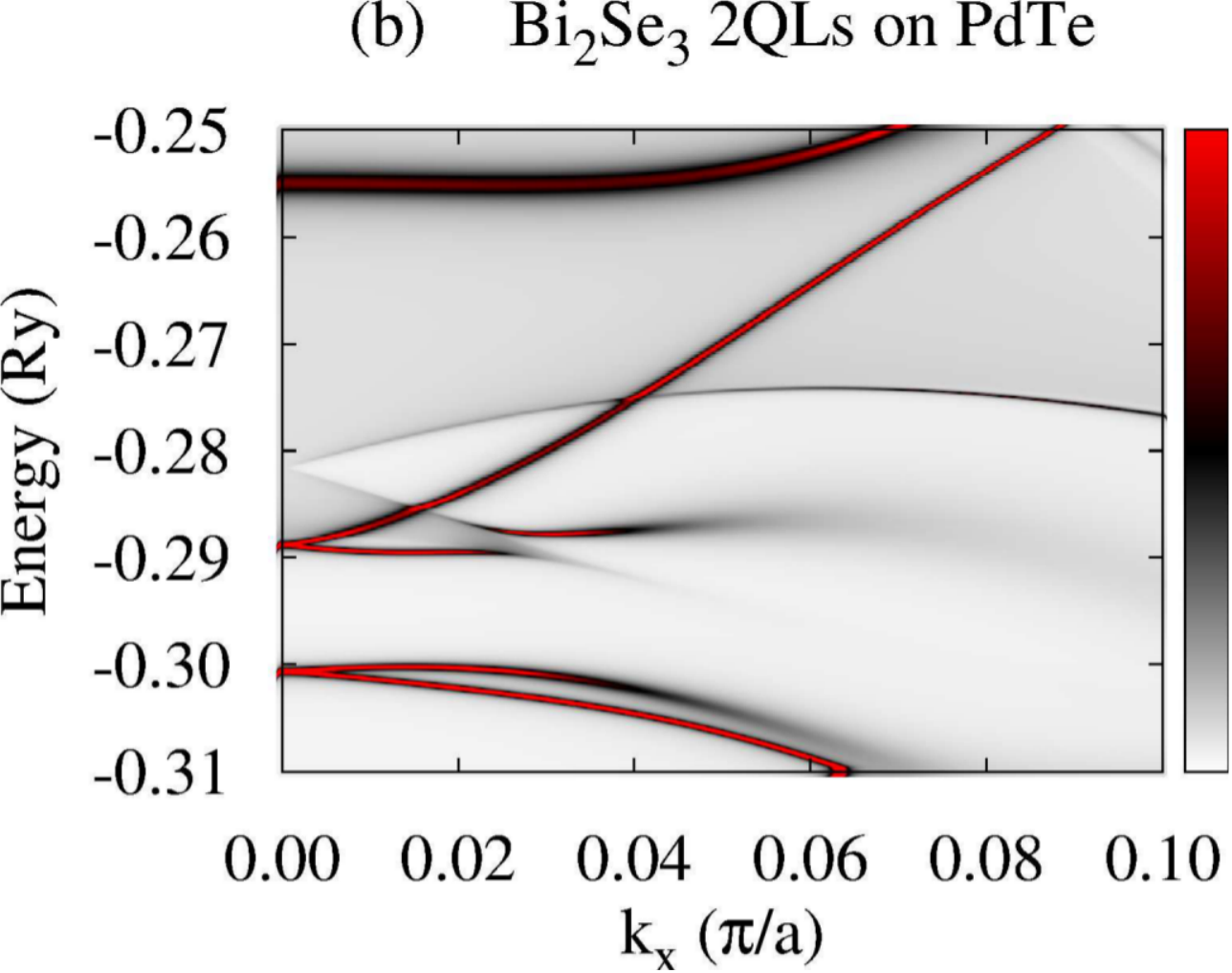}
\hspace{0.2truecm}
\includegraphics[width=0.31\textwidth]{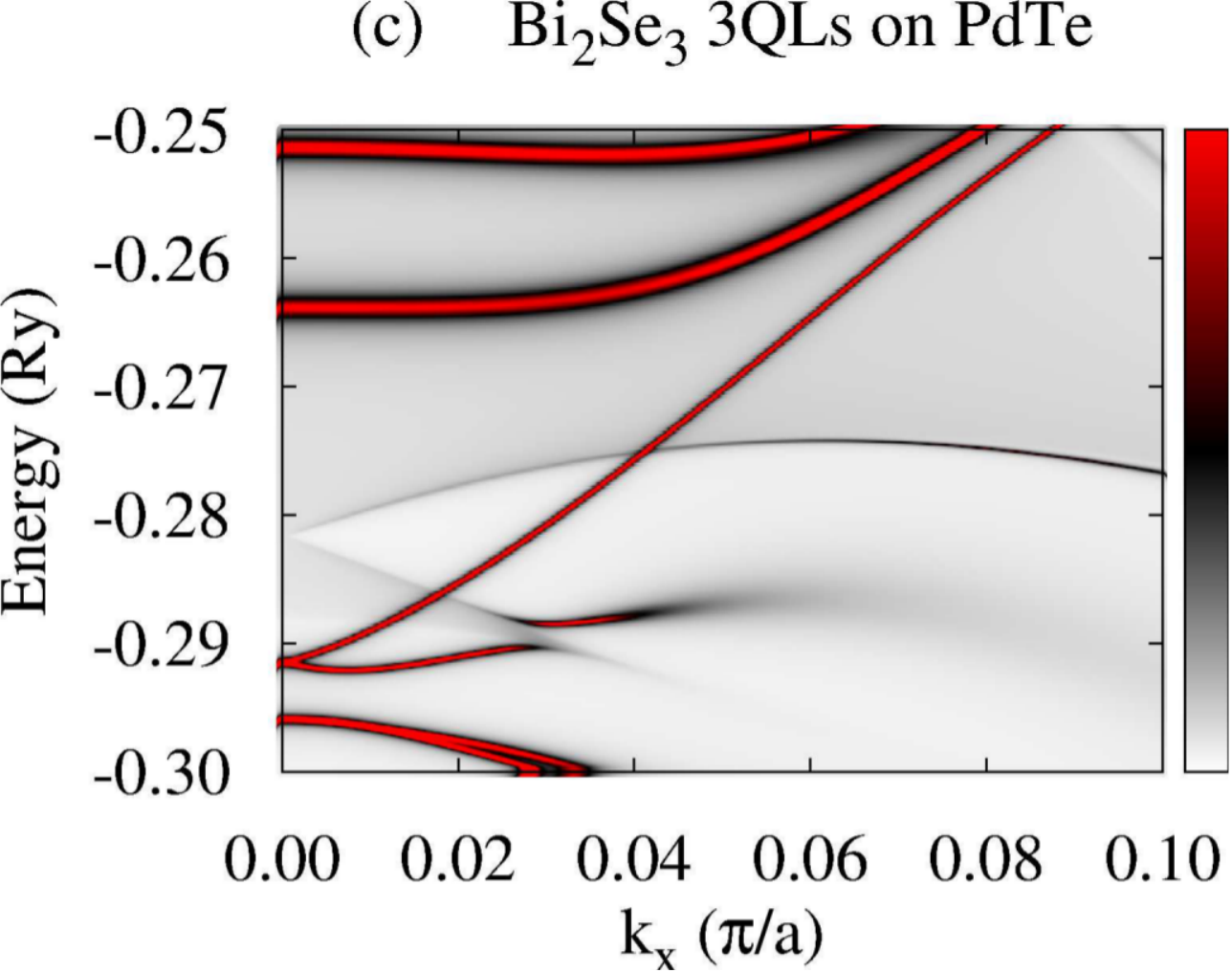}

\vspace{0.2truecm}

\includegraphics[width=0.31\textwidth]{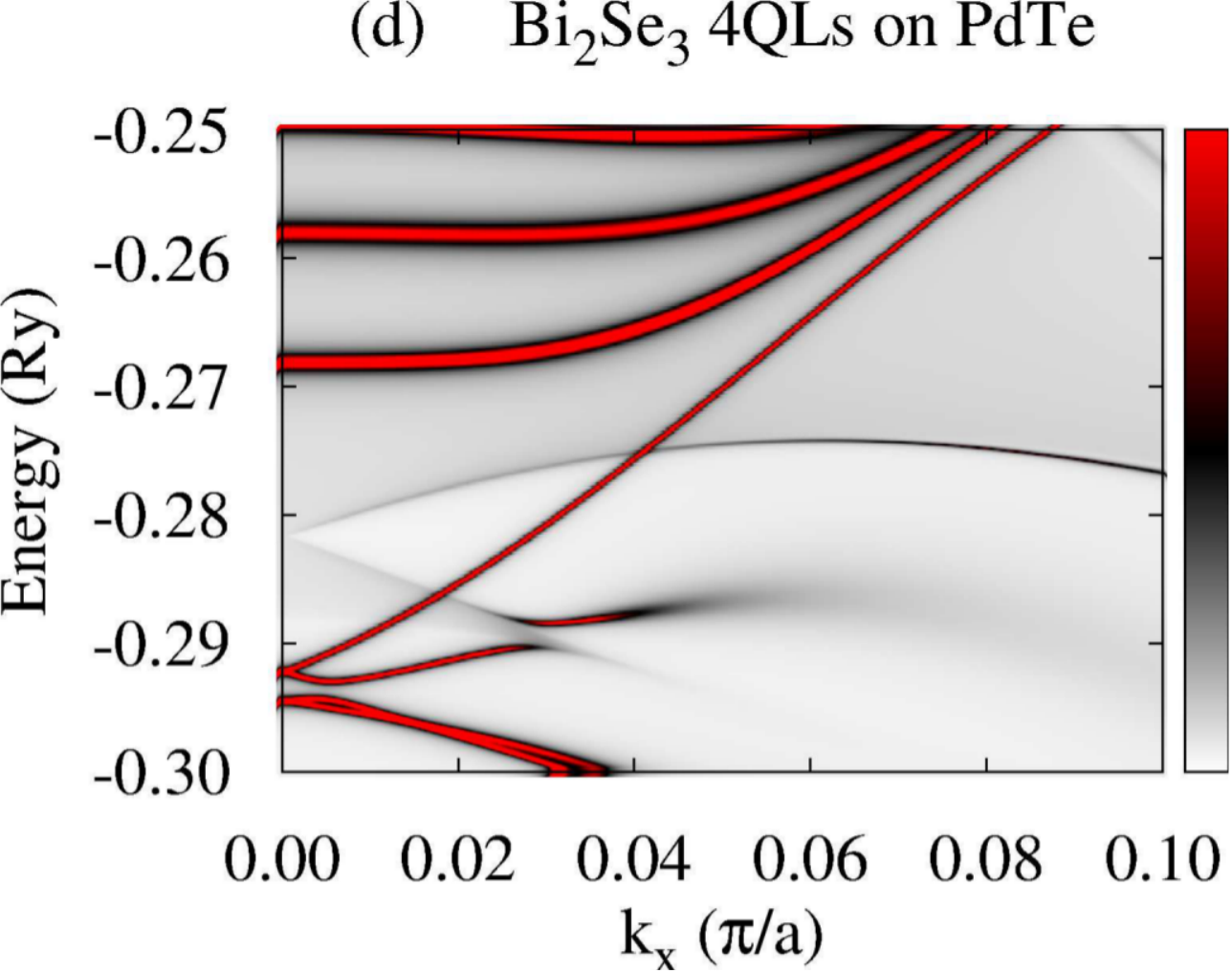}
\hspace{0.2truecm}
\includegraphics[width=0.31\textwidth]{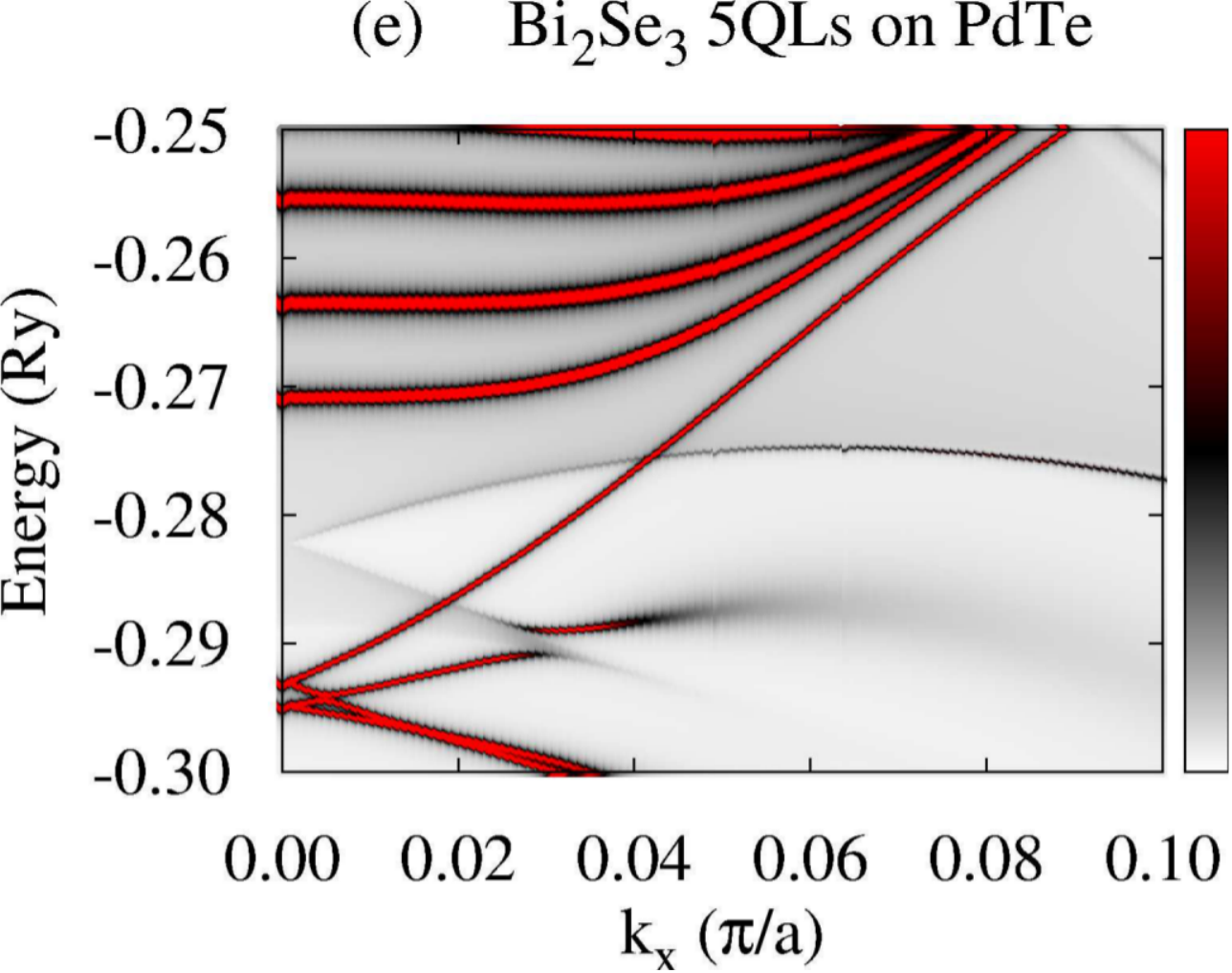}
\hspace{0.2truecm}
\includegraphics[width=0.31\textwidth]{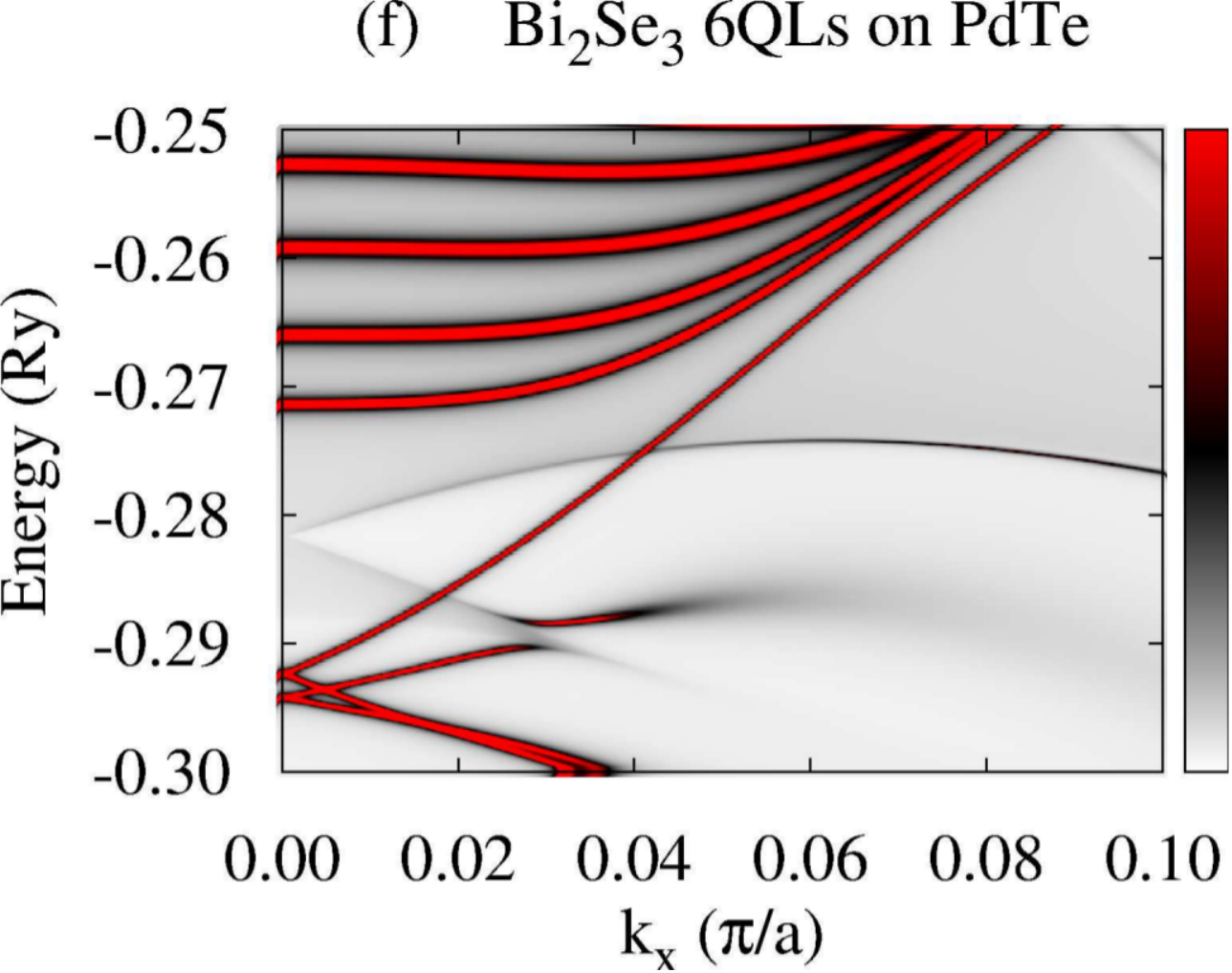}

\vspace{0.2truecm}

\includegraphics[width=0.31\textwidth]{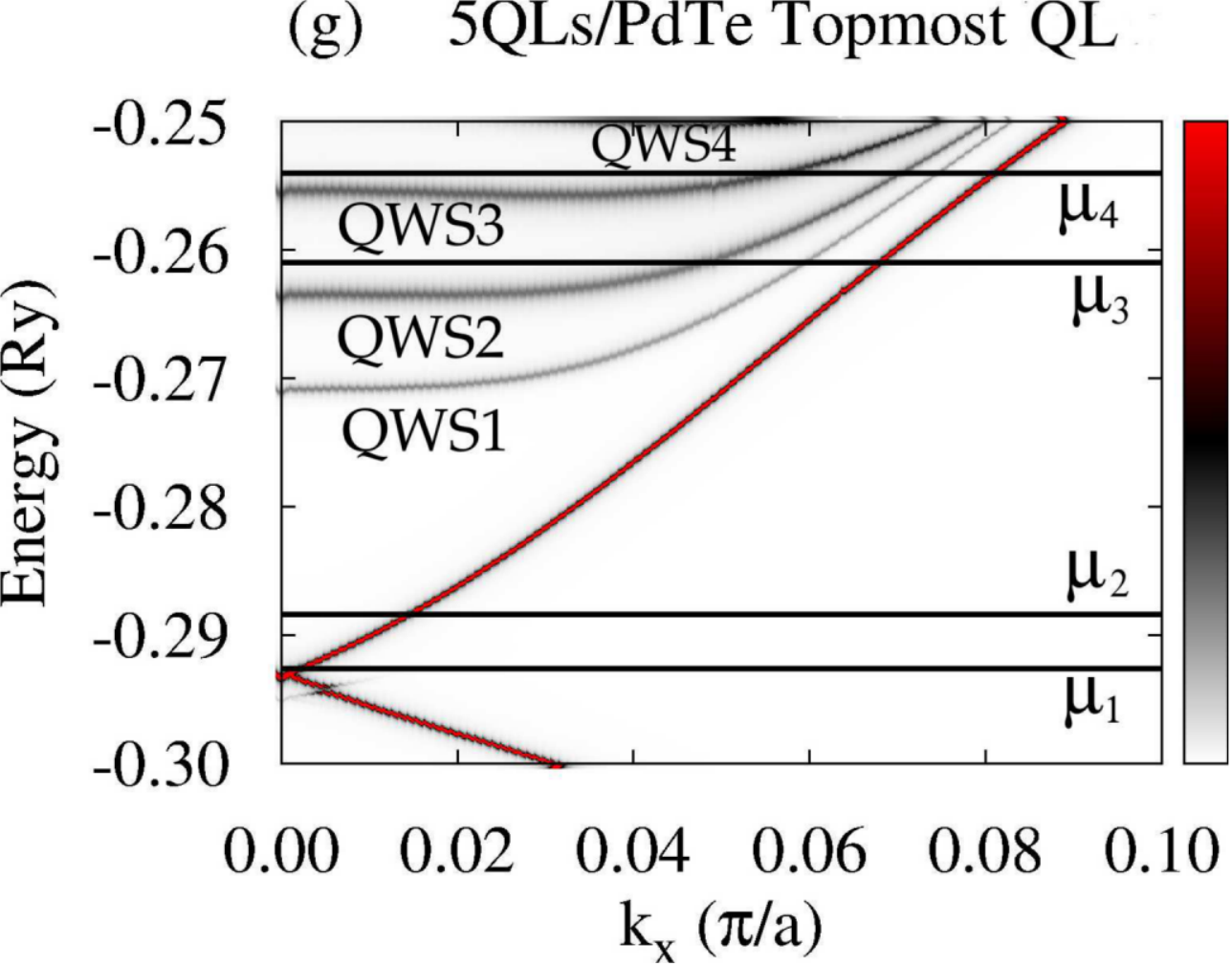}
\hspace{0.2truecm}
\includegraphics[width=0.31\textwidth]{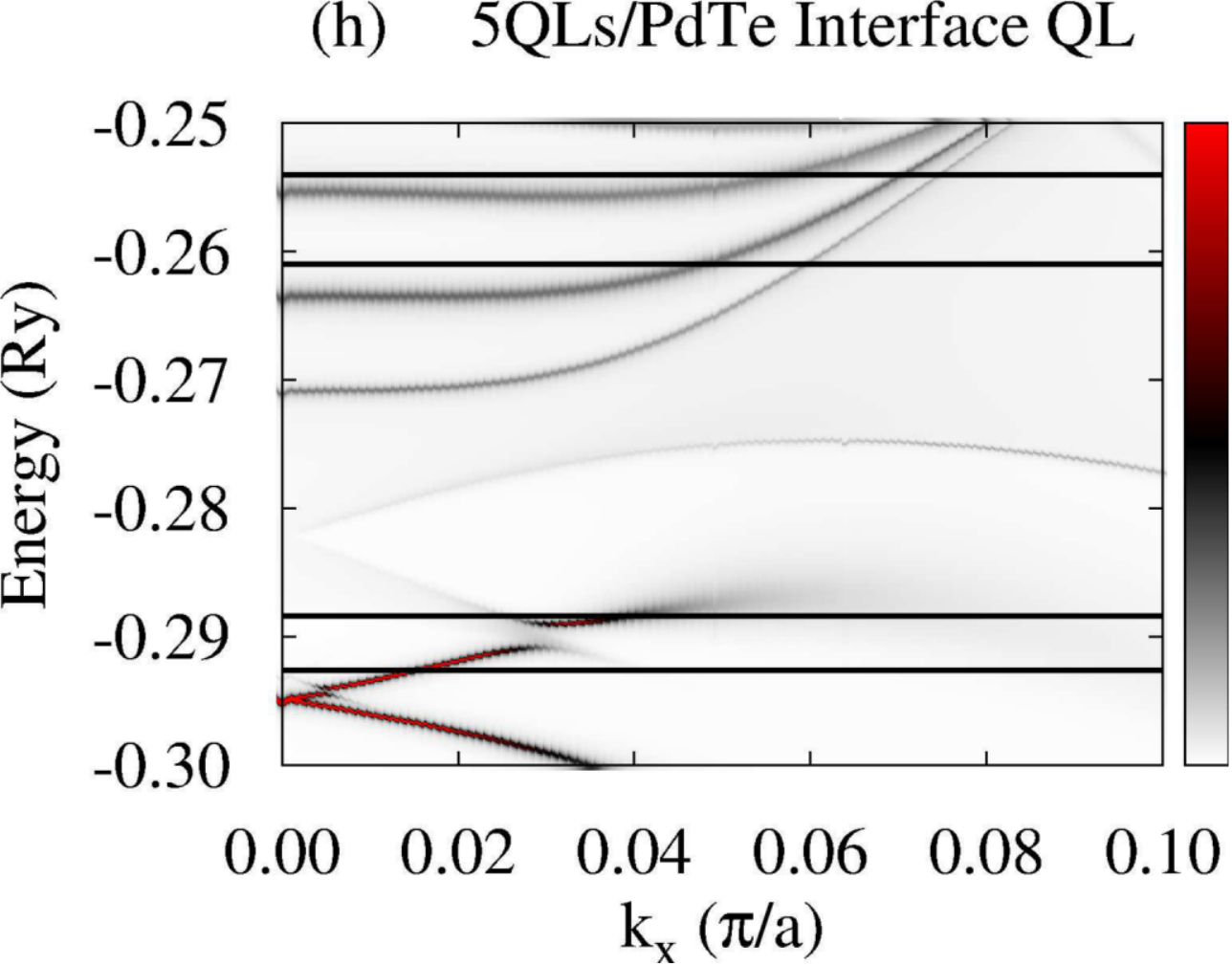}
\hspace{0.2truecm}
\includegraphics[width=0.31\textwidth]{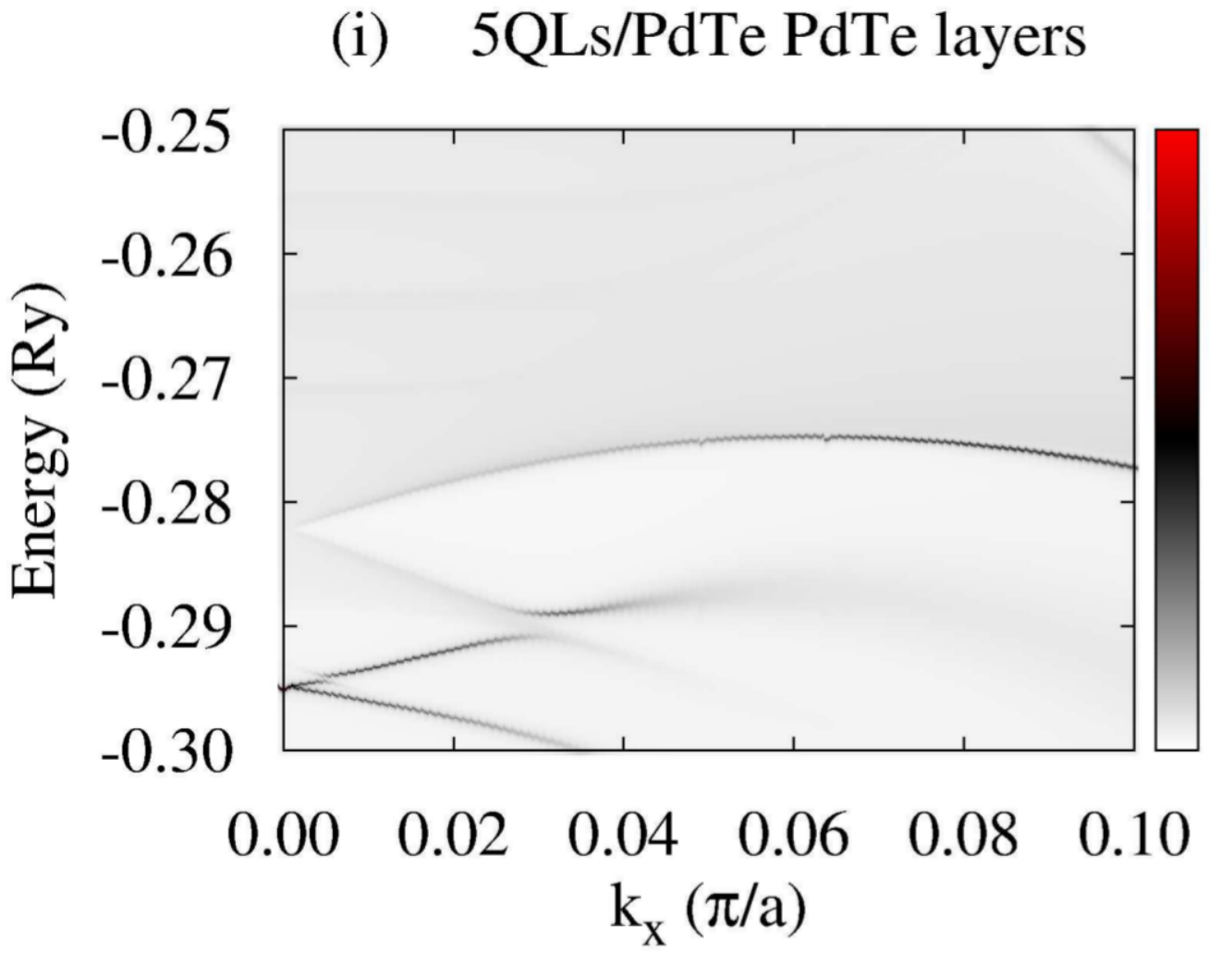}
\caption[Normal_BSF]{(a)-(f) Total BSF contours of the Bi$_2$Se$_3$ overlayers (1-6 QLs) on normal-state PdTe as a function of energy and $k_x$.
BSF contours of (g) the topmost QL, (h) the interface QL, and (i) the PdTe layers for the 5-QL TI overlayer on normal-state PdTe.
The broken inversion symmetry separates the top-surface Dirac states from the interface Dirac states. The surface-hybridization
induced gap appears near $\Gamma$ for (a)-(d). Four horizontal lines in (g) and (h) indicate chemical potential values: $\mu_1=-0.2926$, $\mu_2=-0.2884$, $\mu_3=-0.2610$, and $\mu_4=-0.2540$ Ry. In (g) QWS1, QWS2, QWS3, and QWS4 are four QWS.}
\label{fig:Normal_BSF}
\end{center}
\end{figure*}

First of all, we present calculated electronic structures of PdTe and Bi$_2$Se$_3$ in the normal state, separately. Figure~\ref{fig:geo}(b) and (c)
show BSF contours and DOS of bulk PdTe, respectively. The calculated BSF shows metallic nature. There is a wide range of high electron density
below the Fermi level $E_{\rm F}$ ($-$0.36~Ry) with a sharp DOS peak around $-$1.20~Ry which arises from Te $s$ and $p$ orbitals.
Figure~\ref{fig:geo}(d) shows BSF contours of a semi-infinite Bi$_2$Se$_3$ system. We find Dirac surface states within a bulk band
gap of about 0.03 Ry as well as continuous conduction and valence band regions. For Bi$_2$Se$_3$ slabs, all bands are doubly degenerate due to time-reversal and inversion symmetries, and several QWS appear in the conduction and valence band regions.
For a $N$-QL slab, $N-1$ ($N-2$) QWS appear in the conduction (valence) band region \cite{Park2013}, as shown in
Fig.~S1(b),(d) in the SM. For Bi$_2$Se$_3$ slabs thinner than 5-6 QLs, the top and bottom surface states hybridize, opening
an energy gap in the Dirac surface states \cite{Park2013,YZhang2010}. This gap is referred to as a surface-hybridization gap.
The SKKR-calculated band structures of PdTe and Bi$_2$Se$_3$ agree with those using VASP code \cite{VASP1,VASP2}
(Fig.~S1 in the SM).

%\subsection{Bi$_2$Se$_3$ films on PdTe}

Using the above converged potentials of PdTe and Bi$_2$Se$_3$, we perform fully relativistic SKKR calculations on the
Bi$_2$Se$_3$-PdTe heterostructure in the normal state. Figure~\ref{fig:Normal_BSF}(a)-(f) show calculated normal-state BSF of the heterostructure
with the TI film thickness varying from 1 to 6 QLs, where all layers are summed. The gray continuous spectrum in the BSF is similar to that of PdTe.
Compare Fig.~\ref{fig:Normal_BSF}(a)-(f) with Fig.~\ref{fig:geo}(b) or \ref{fig:Normal_BSF}(i). We find that a shift of the Madelung potential
lowers the TI Dirac point around $-0.29$~Ry and that the top-surface and interface Dirac
states are shifted from each other with strong modification of the dispersion of the interface Dirac states. The slope of the dispersion near the
Dirac point is substantially reduced and the states lose the interface-state character somewhat away from $\Gamma$.
See Fig.~\ref{fig:Normal_BSF}(g) and (h). The top-surface (interface) Dirac states are identified as states with a large BSF weight onto the
topmost (interface) QL. Strong hybridization of the interface states with the substrate causes the strong modification of their dispersion.
Similar effects have been reported in various heterostructures involving Bi$_2$Se$_3$ \cite{Govaerts2014,Hsu2017}.
For thin films ($<$~5 QLs) we also observe an energy gap in the vicinity of the two Dirac points which decreases with increasing the TI overlayer
thickness. This gap is induced by the hybridization between the interface and top-surface Dirac states. The shape and number of TI QWS, however,
remain unchanged with the substrate, but they are quite broadened compared to the Dirac states, as shown in Fig.~\ref{fig:Normal_BSF}(b)-(f).
This broadening may be caused by scattering of electrons from the substrate. Calculated BSF with finer resolution suggests that each broad QWS
peak in Fig.~\ref{fig:Normal_BSF} consists of two bands or states. For the 5-QL overlayer on PdTe [Fig.~\ref{fig:Normal_BSF}(g)],
chemical potential $\mu_4=-0.2540$~Ry crosses three QWS (labeled as QWS1, QWS2, QWS3) and the top-surface Dirac state, while chemical potential $\mu_2=-0.2844$~Ry crosses only the top-surface and interface Dirac states. For thinner TI overlayers, chemical potential $\mu_4$ crosses fewer
number of QWS compared to the 5-QL overlayer. See Fig.~S2 for the characteristics of the QWS.

%%%%%%%%%%%%%%%%%%%%%%%%%%%%%%%%%%%%%%%%%%%%%%%%%%%%%%%%%%%%%%%%%%%%%%%%%%%

\subsection{Electronic structure of superconducting state}

\subsubsection{Induced spectral gap with $\Delta^{\rm{TI}}_{\rm{eff}}=0$}\label{sec3:gap1}

\begin{figure*}[!h]
\begin{center}
\includegraphics[width=0.31\textwidth]{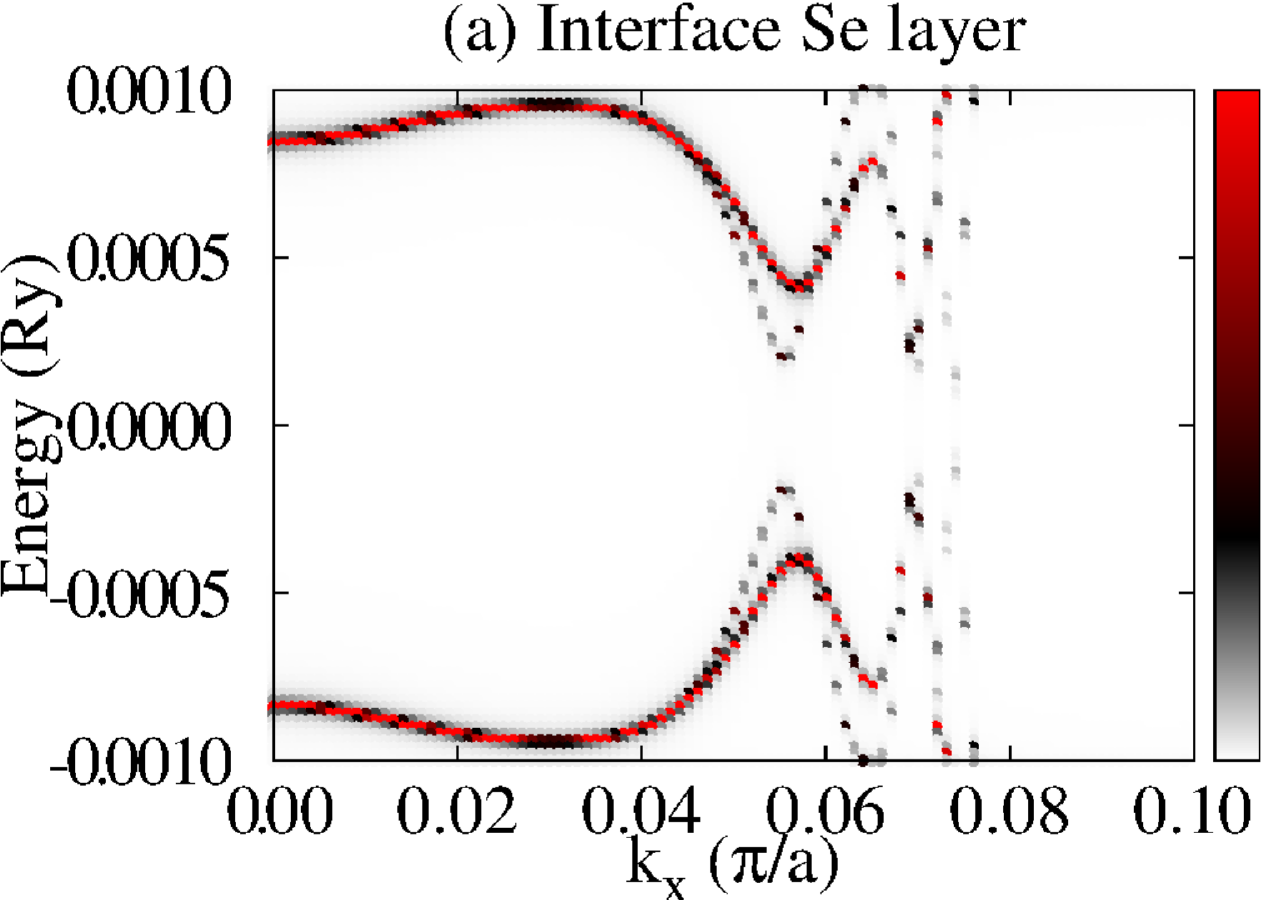}
\hspace{0.2truecm}
\includegraphics[width=0.31\textwidth]{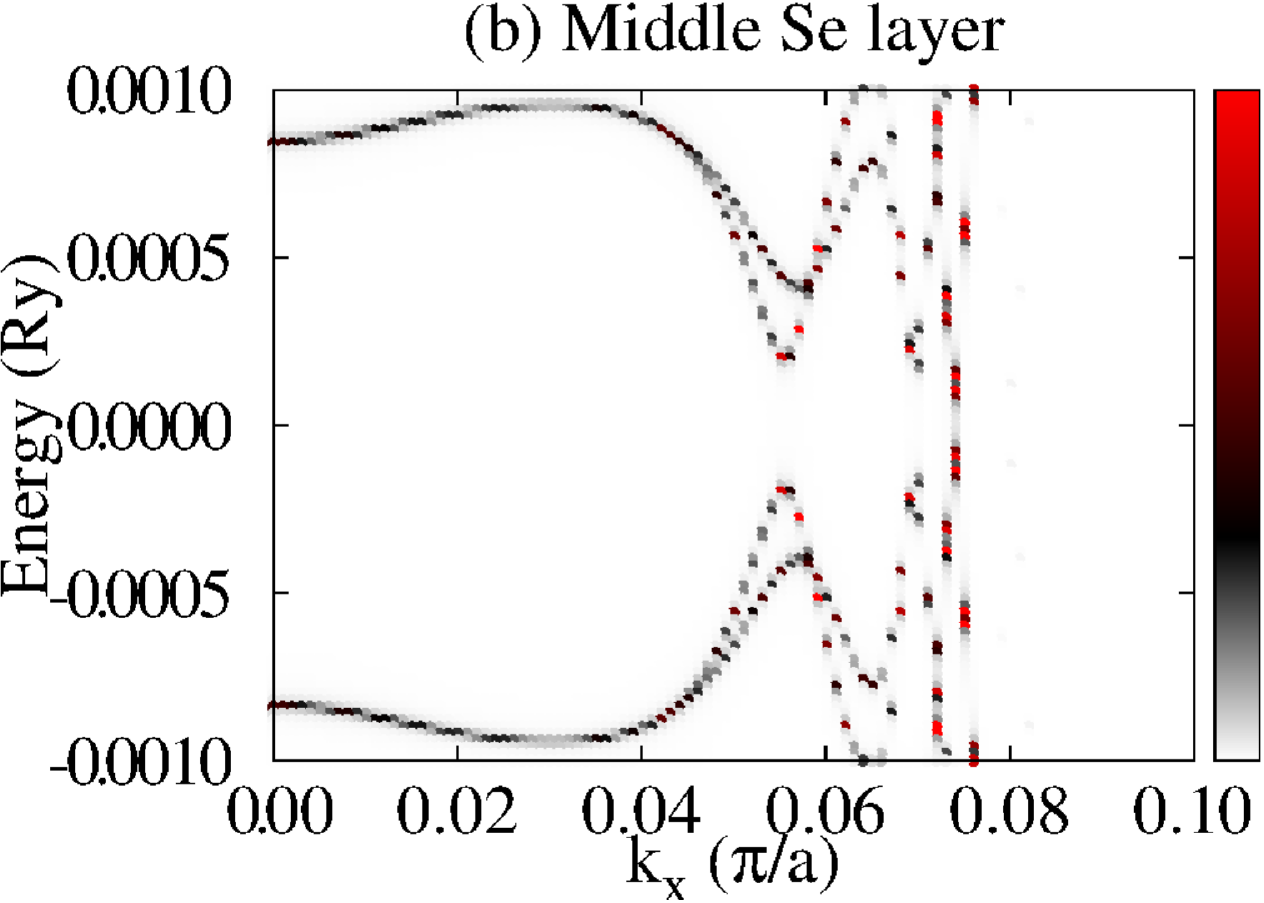}
\hspace{0.2truecm}
\includegraphics[width=0.31\textwidth]{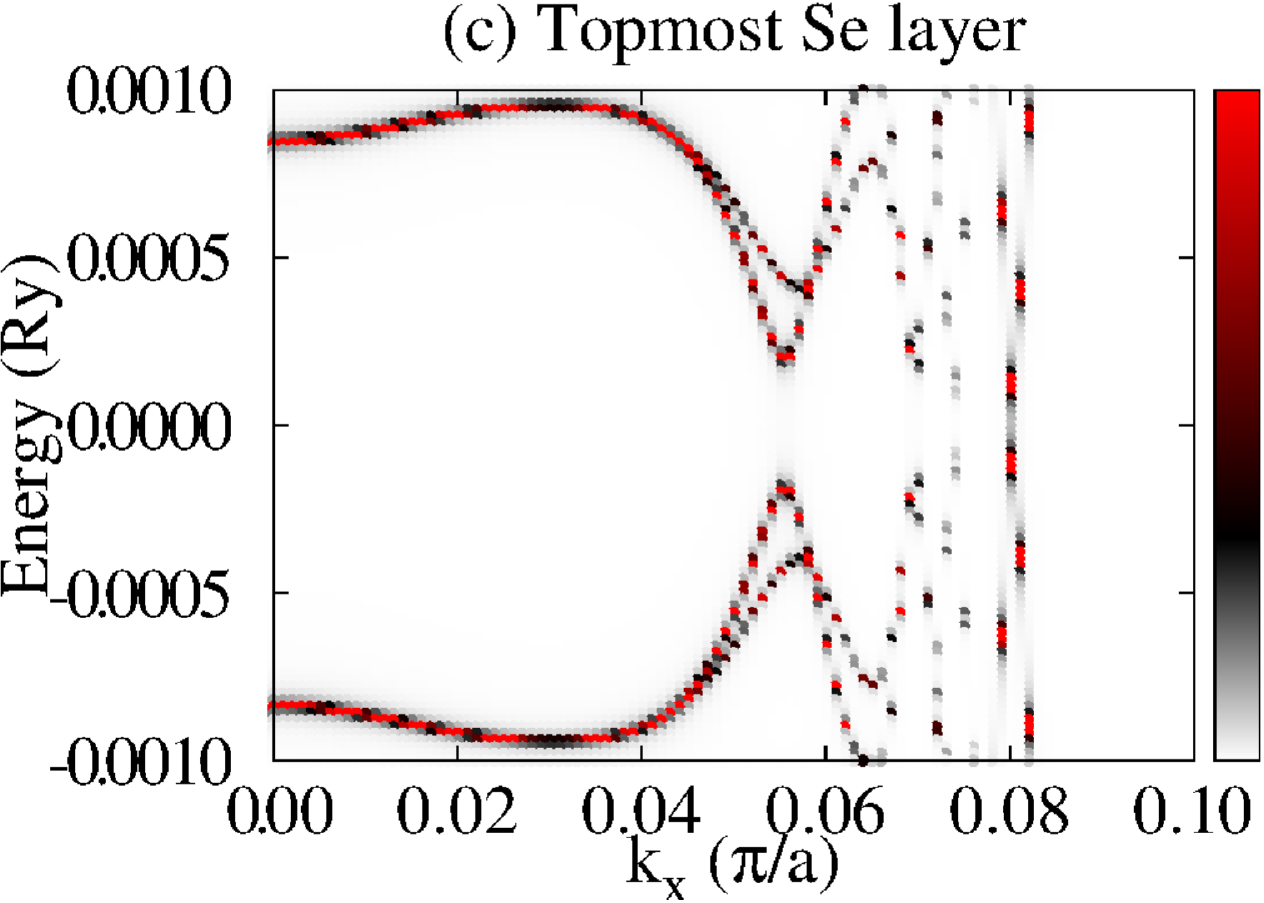}

\vspace{0.2truecm}

\includegraphics[width=0.31\textwidth]{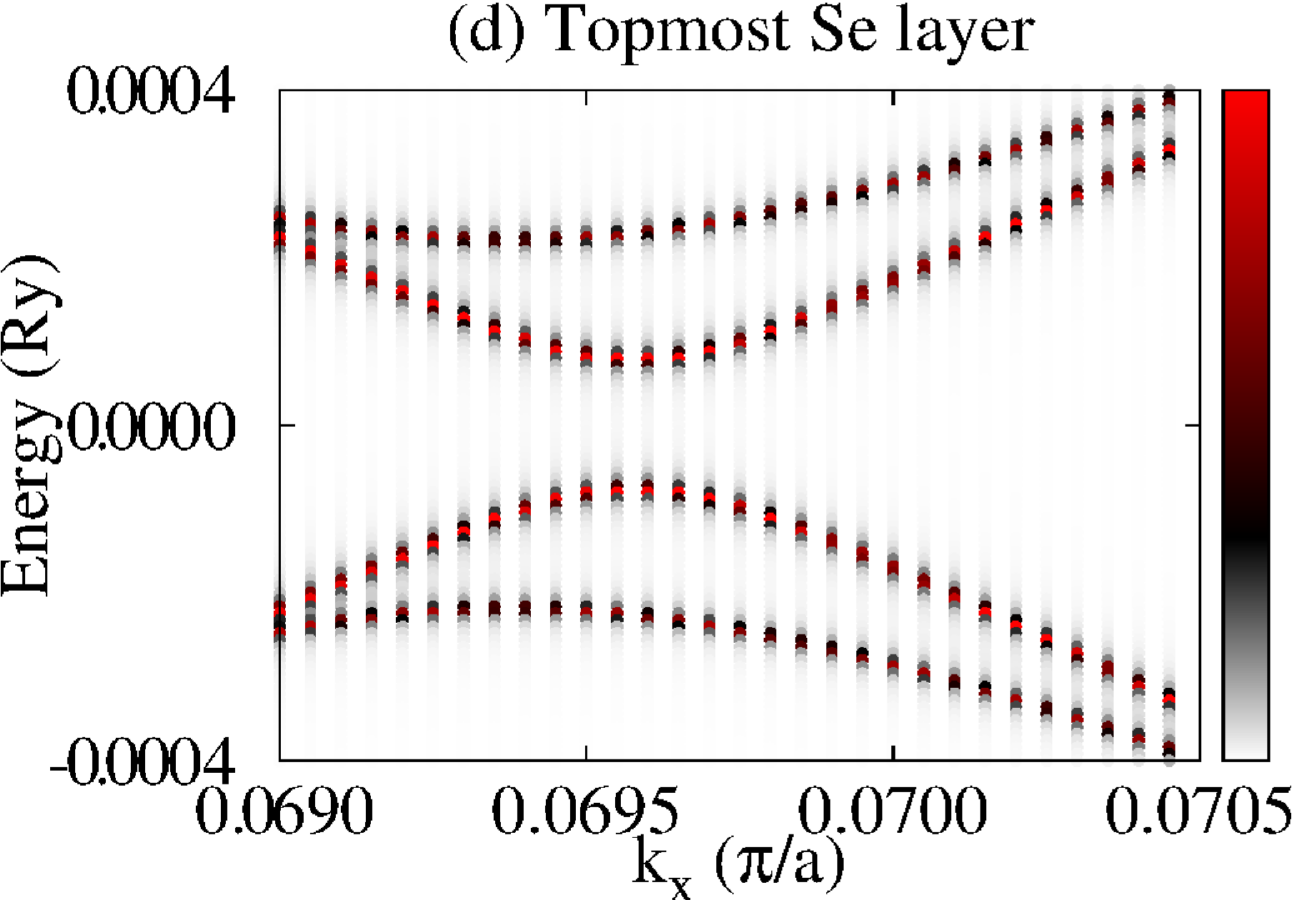}
\hspace{0.2truecm}
\includegraphics[width=0.31\textwidth]{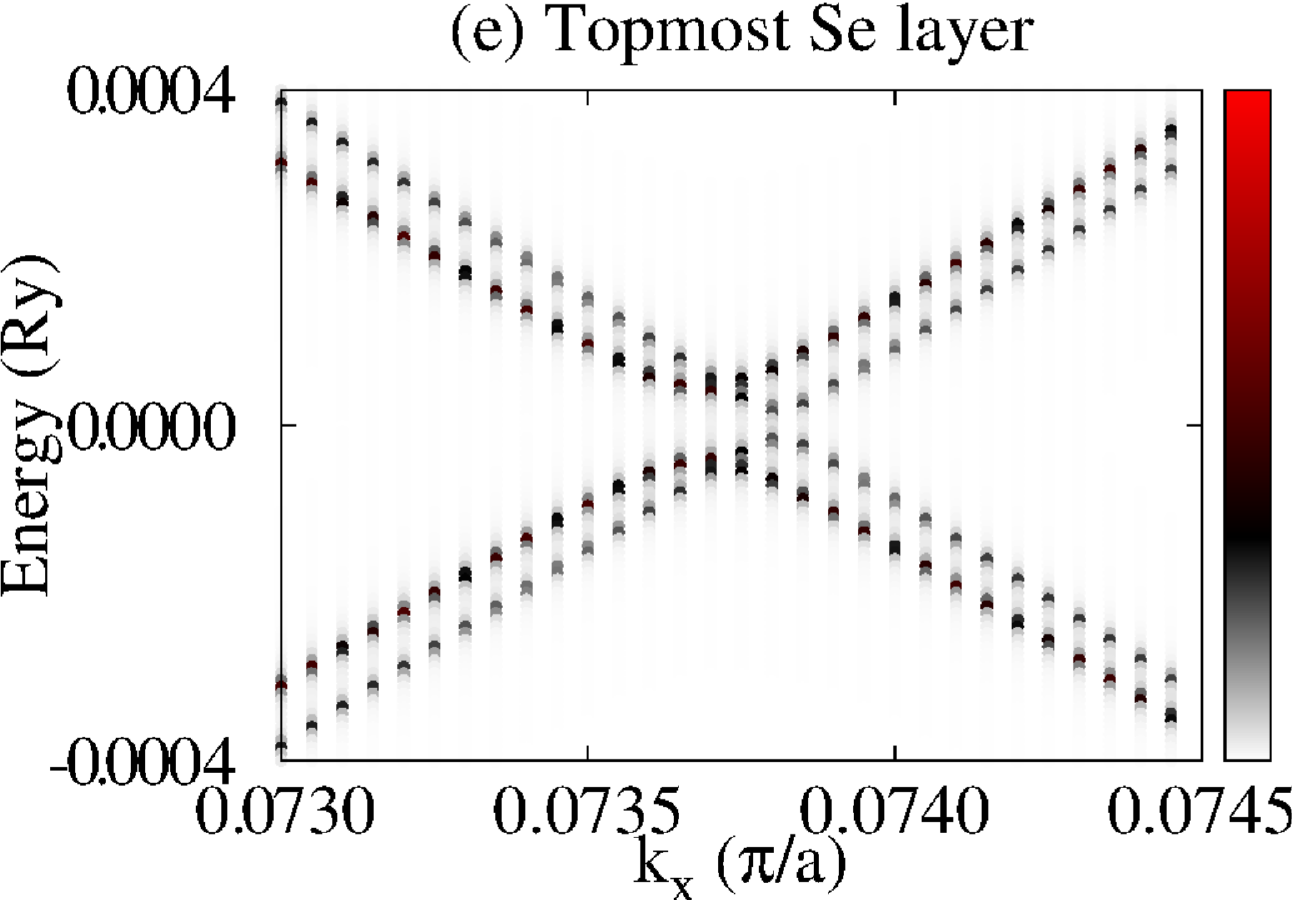}
\hspace{0.2truecm}
\includegraphics[width=0.31\textwidth]{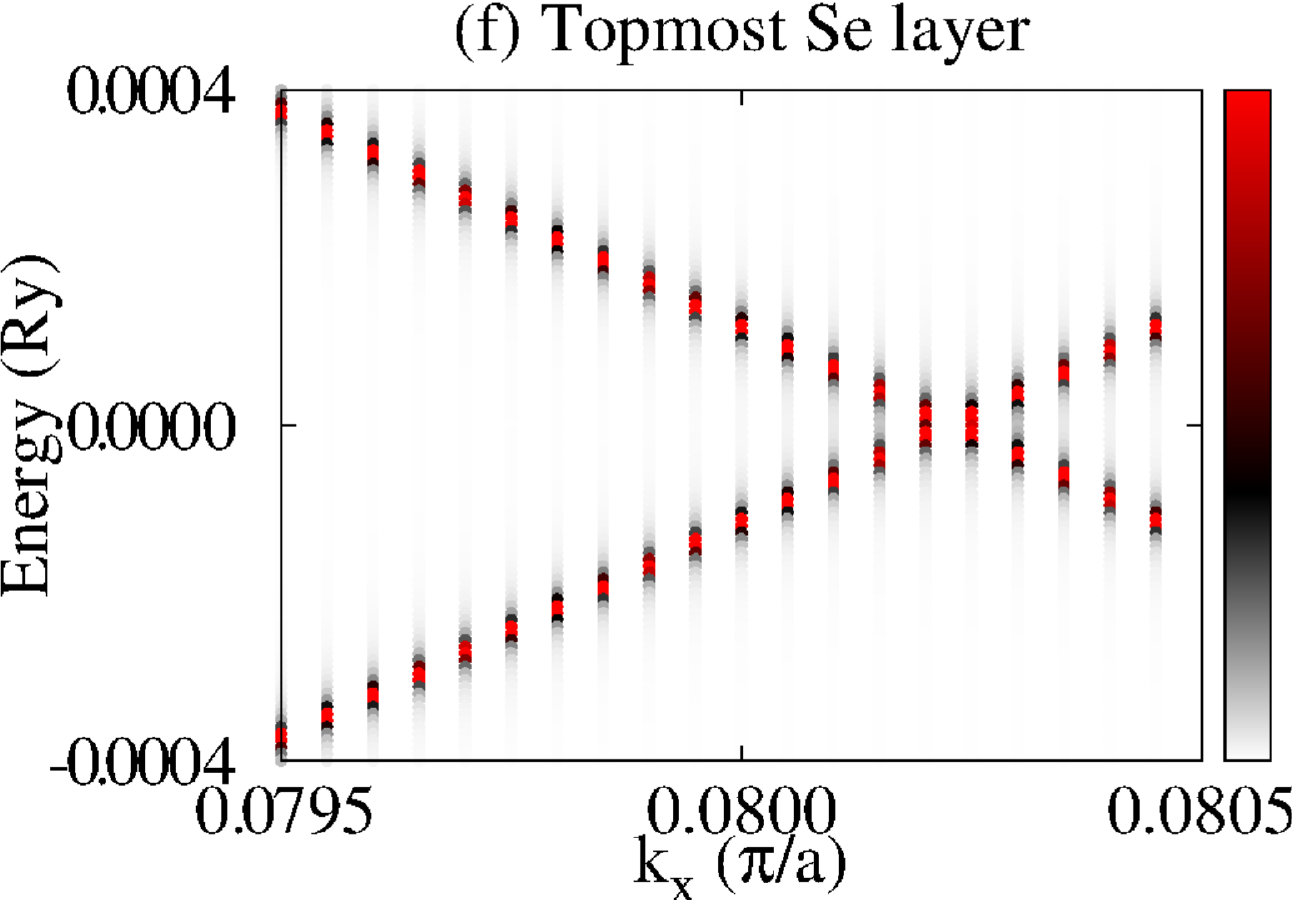}

\vspace{0.2truecm}

\includegraphics[width=0.31\textwidth]{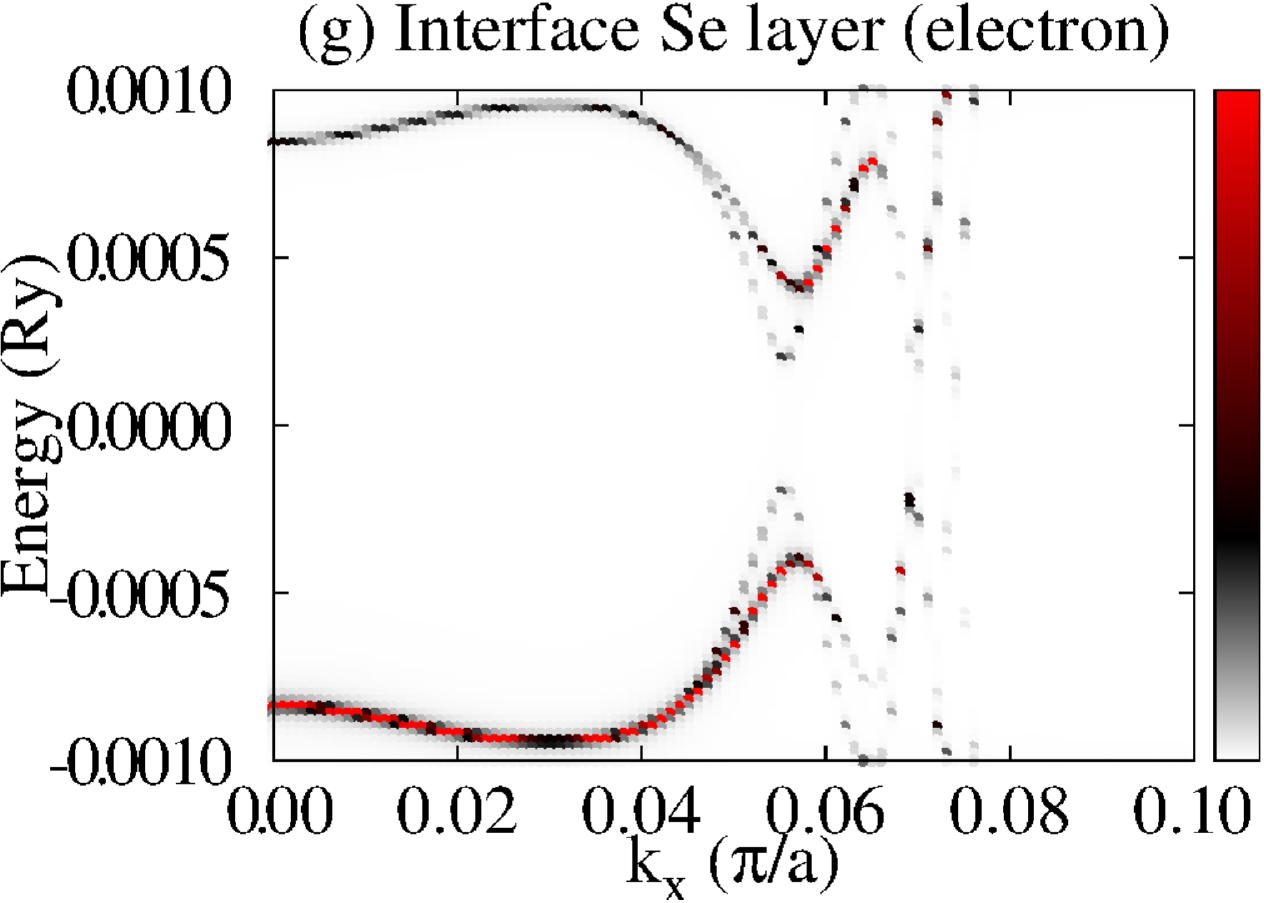}
\hspace{0.2truecm}
\includegraphics[width=0.31\textwidth]{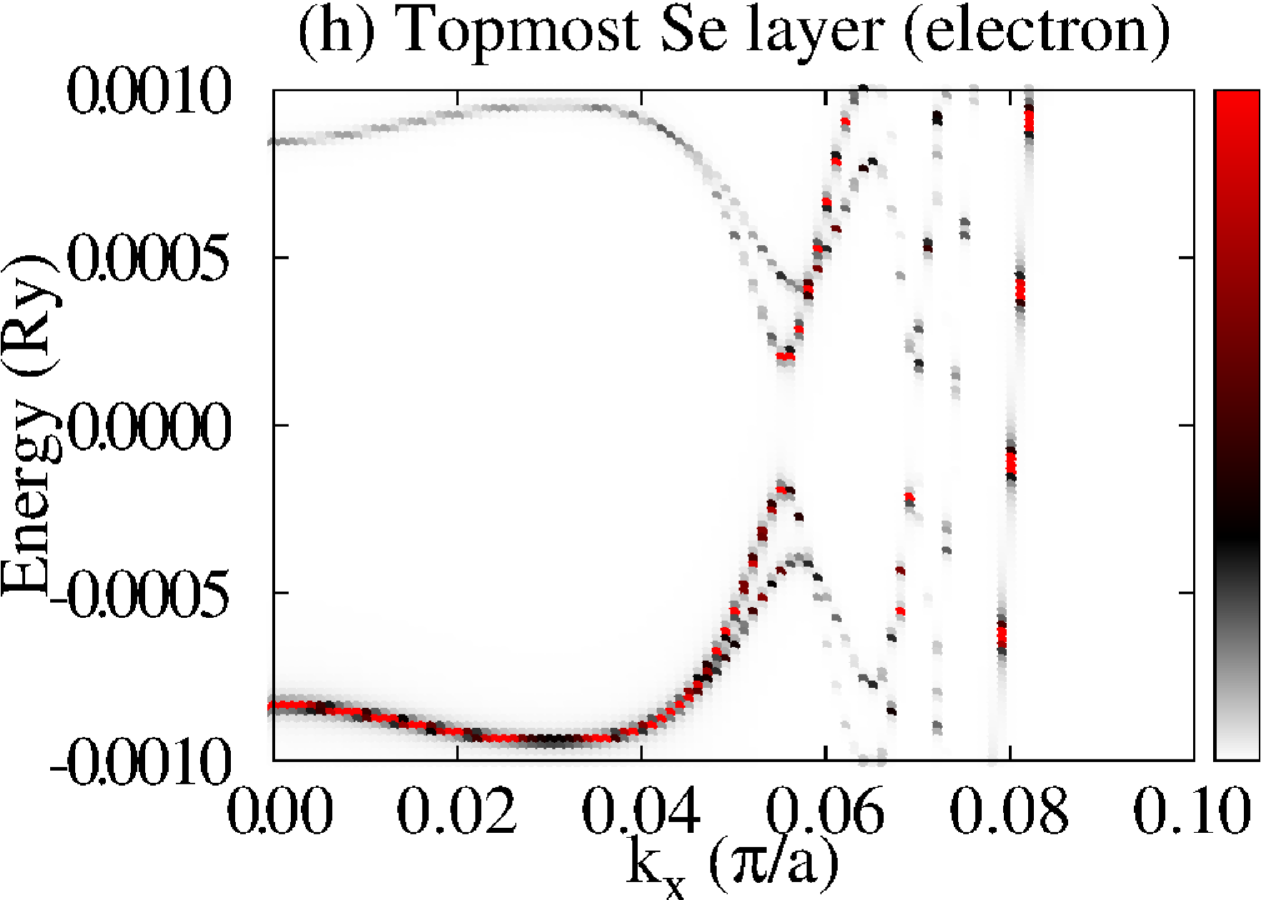}
\hspace{0.2truecm}
\includegraphics[width=0.31\textwidth]{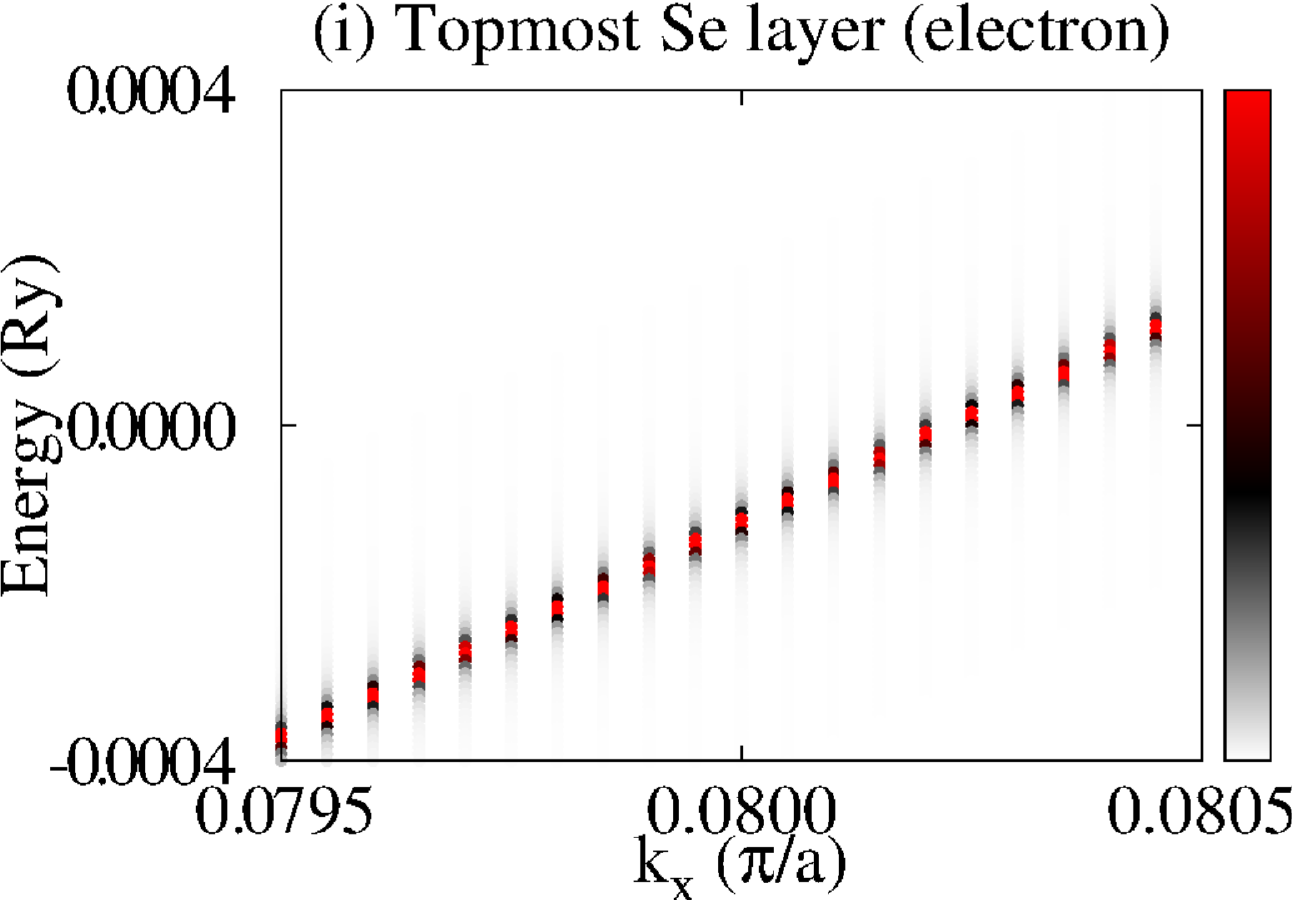}

\vspace{0.2truecm}

\includegraphics[width=0.31\textwidth]{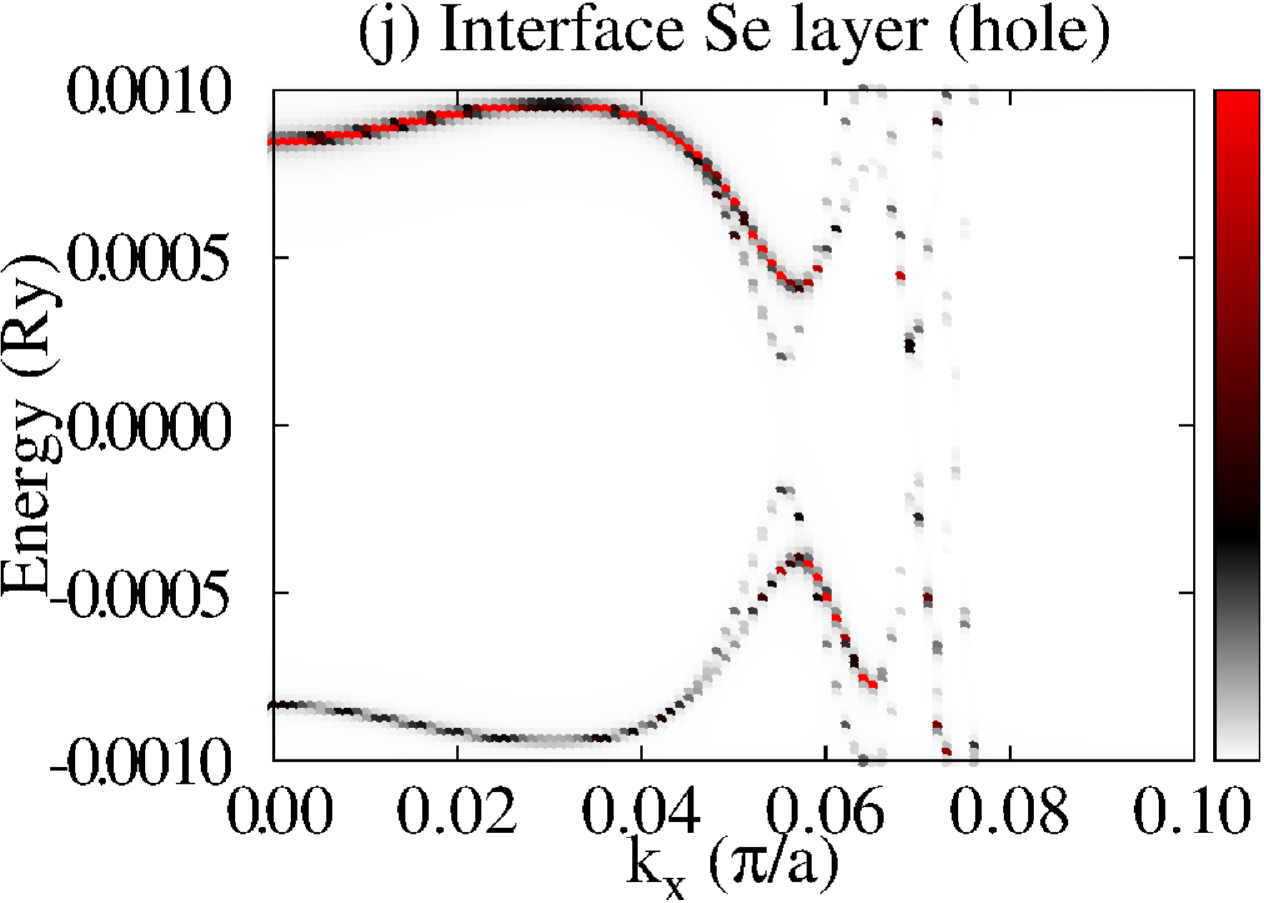}
\hspace{0.2truecm}
\includegraphics[width=0.31\textwidth]{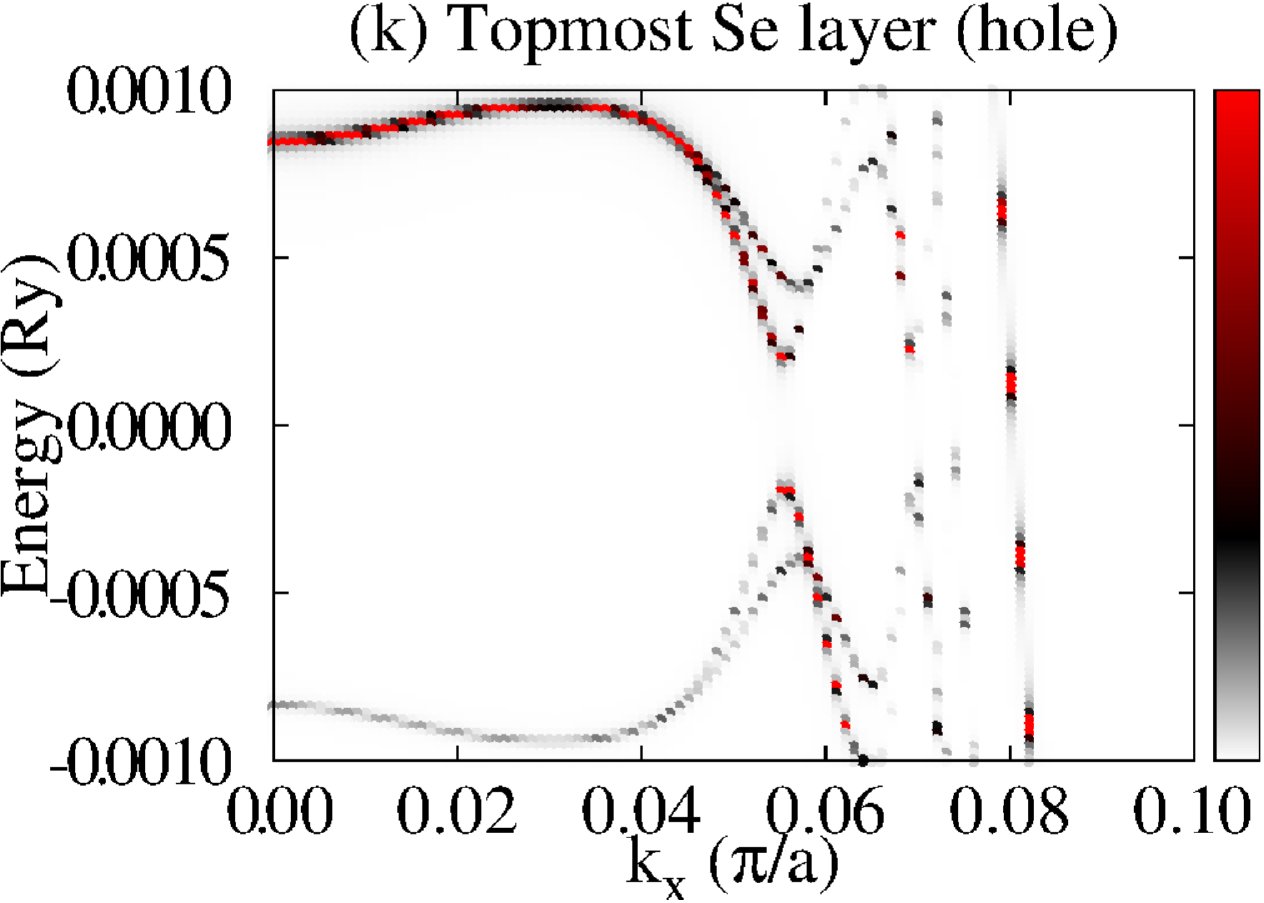}
\hspace{0.2truecm}
\includegraphics[width=0.31\textwidth]{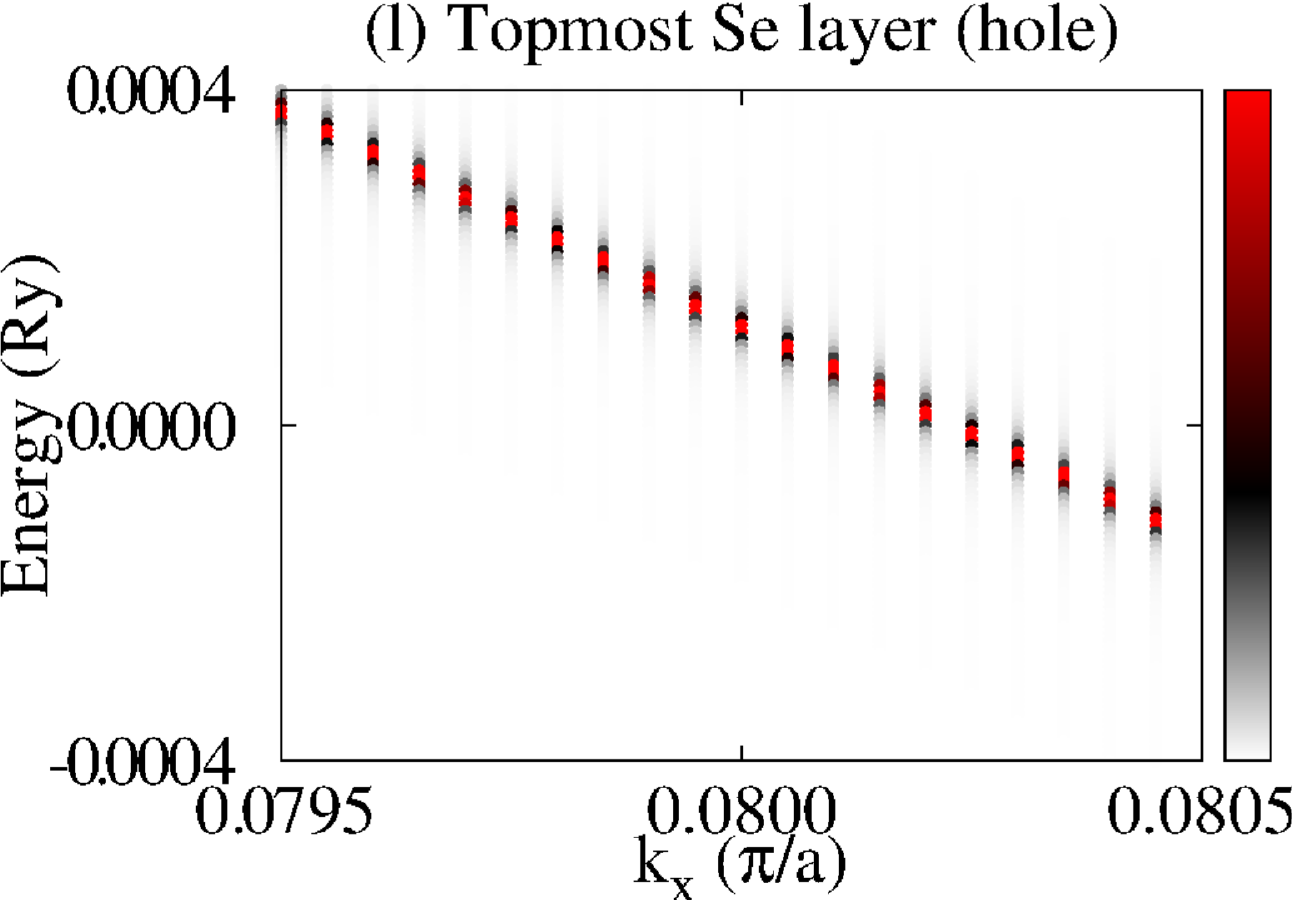}
\caption[SC_BSF]{BSF contours of the 5-QL overlayer on {\it SC} PdTe at $\mu_4=-0.2540$ Ry.
(a) Interface Se, (b) middle Se, and (c) topmost Se layer BSF.
%In (c) four types of proximity-induced gap are observed near $k_x=0.05550$, 0.06950, 0.07350, and 0.08023~$\pi/a$.
(d)-(f) Zoom-in of (c) near the induced gaps $\Delta_{\bf k3}$, $\Delta_{\bf k4}$, $\Delta_{\bf k5}$, $\Delta_{\bf k6}$, and $\Delta_{\bf k7}$.
In (a)-(f) both electron and hole contributions are included. (g),(j) Electron and hole BSF contours of (a).
(h),(k) Electron and hole BSF contours of (c). (i),(l) Electron and hole BSF contours of (f).
The smallest induced gap $\Delta_{\bf k7}$ is less than 0.1\% of the bulk SC gap, as shown in (f).}
\label{fig:SC_BSF}
\end{center}
\end{figure*}

%1) Feature of the induced gap at a fixed thickness and fixed chemical potential mu_4
%2) Dependence of the gaps on the chemical potential at a fixed overlayer thickness.
%3) Dependence of the gaps on the overlayer thickness.

\begin{figure*}[!h]
\begin{center}
\includegraphics[width=0.28\textwidth]{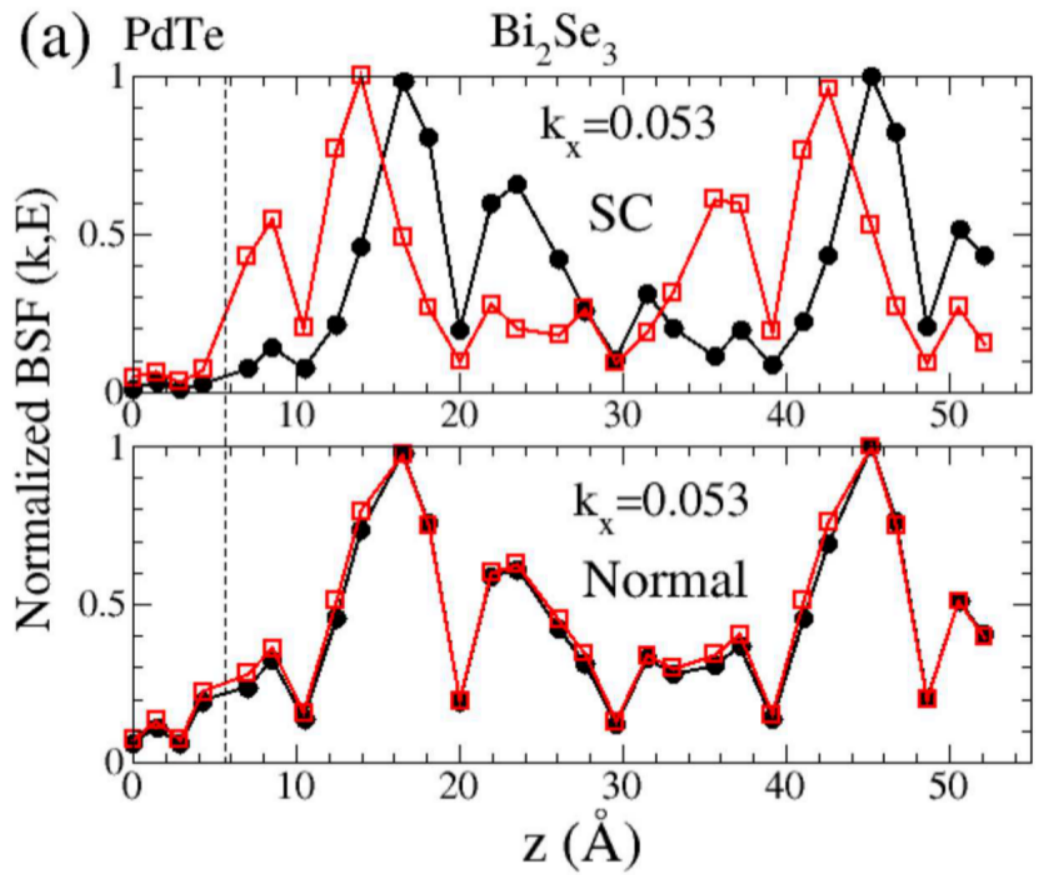}
\hspace{0.2truecm}
\includegraphics[width=0.29\textwidth]{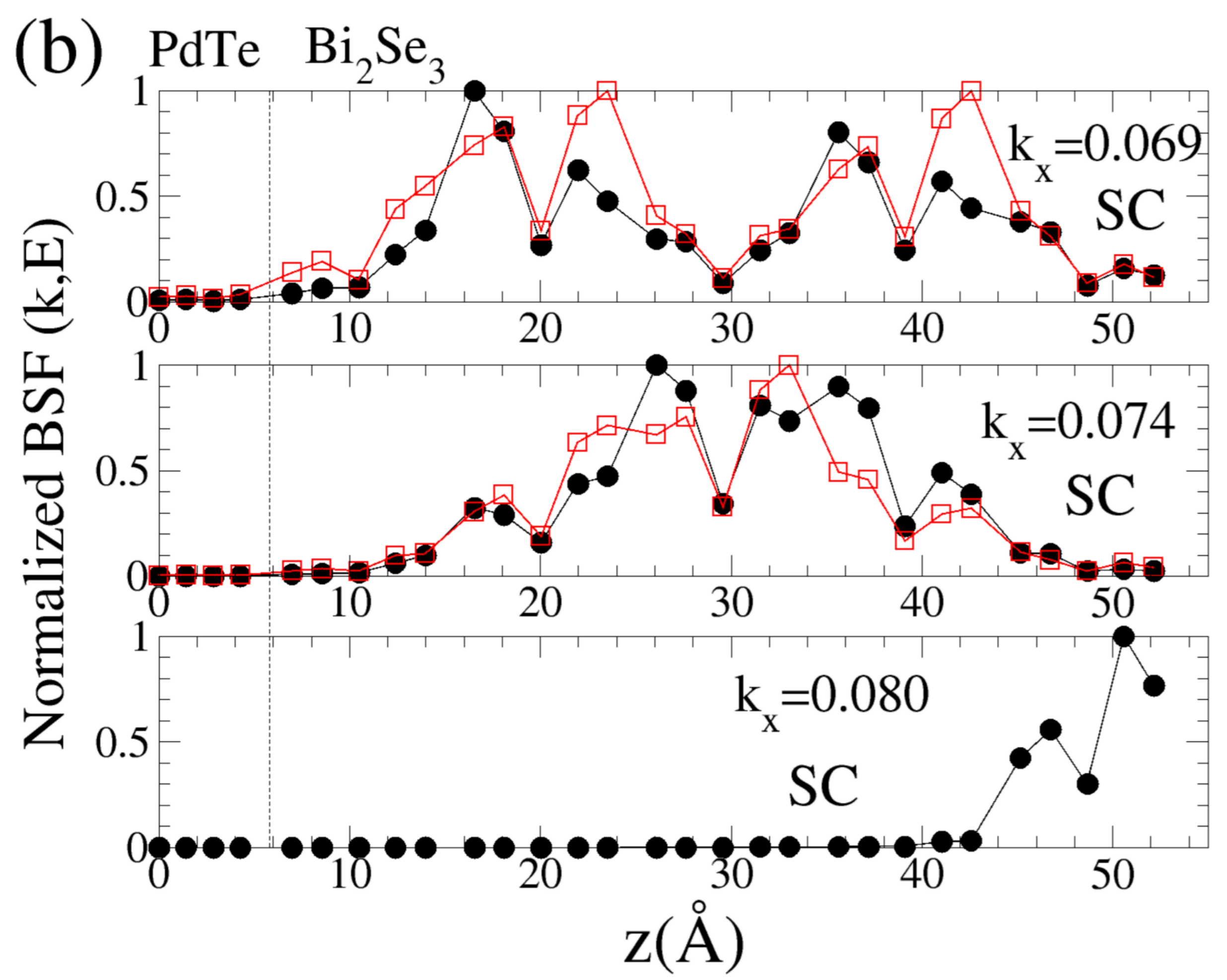}
\hspace{0.2truecm}
\includegraphics[width=0.38\textwidth]{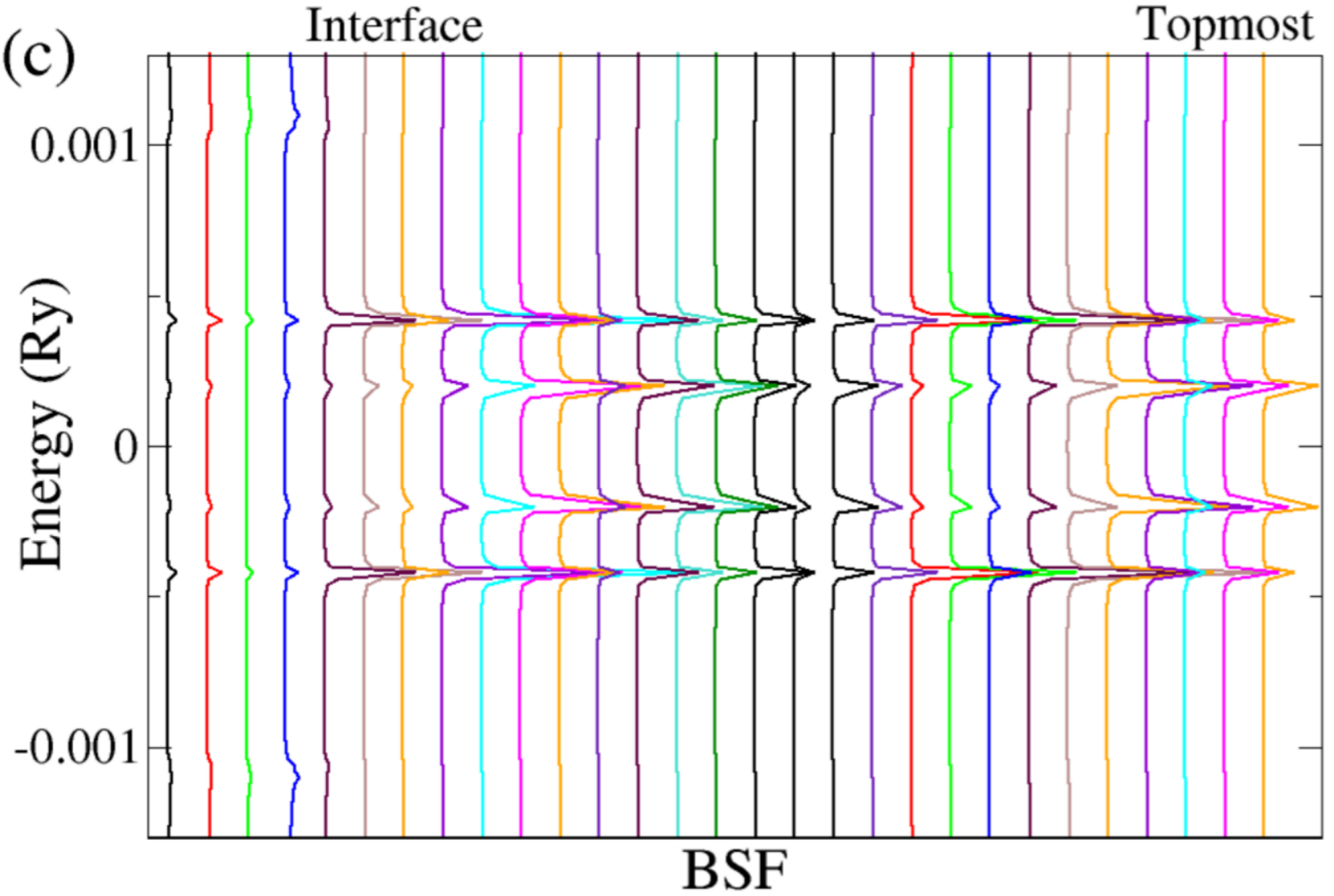}

\vspace{0.2truecm}

\includegraphics[width=0.38\textwidth]{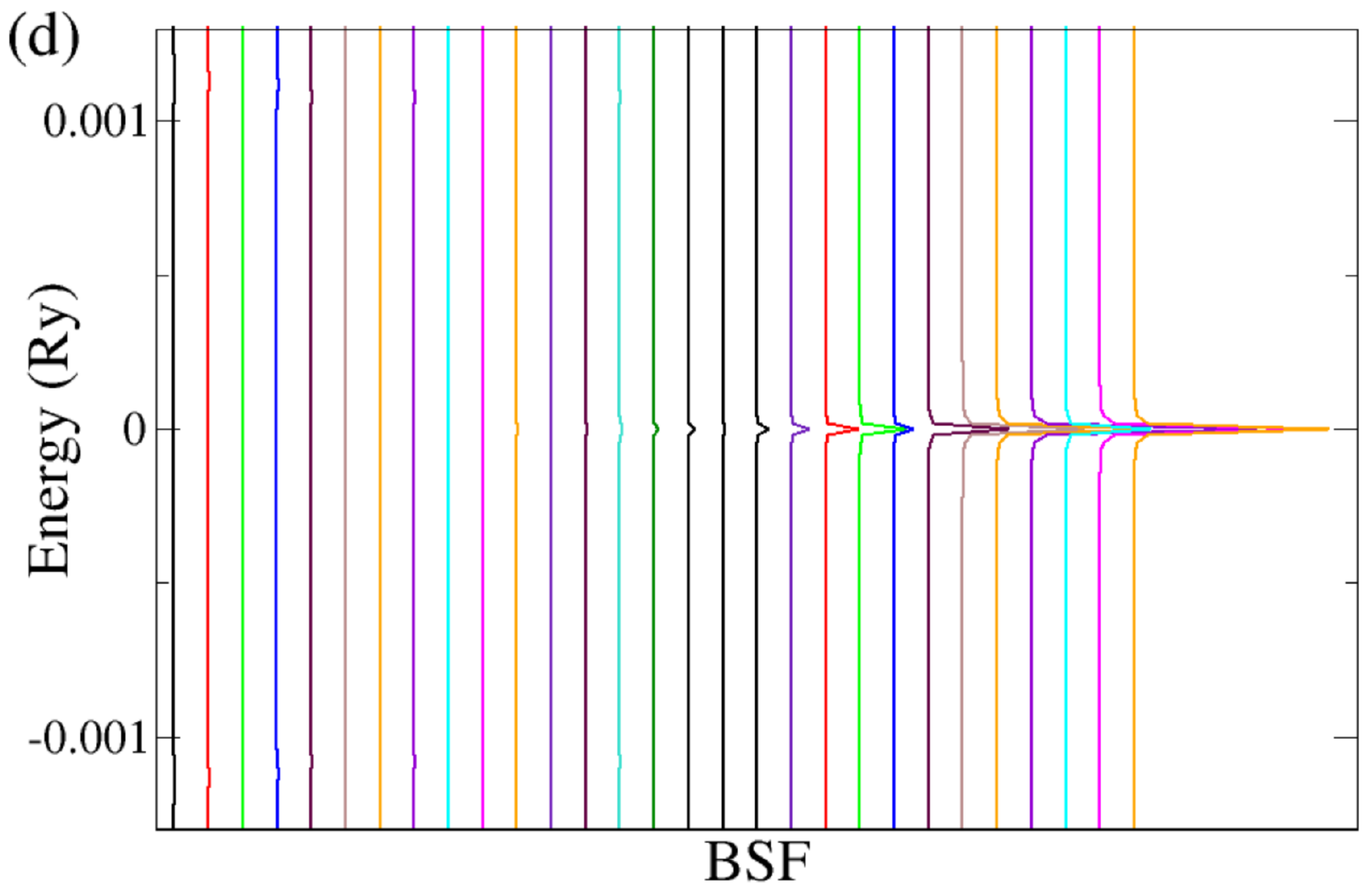}
\hspace{0.2truecm}
\includegraphics[width=0.38\textwidth]{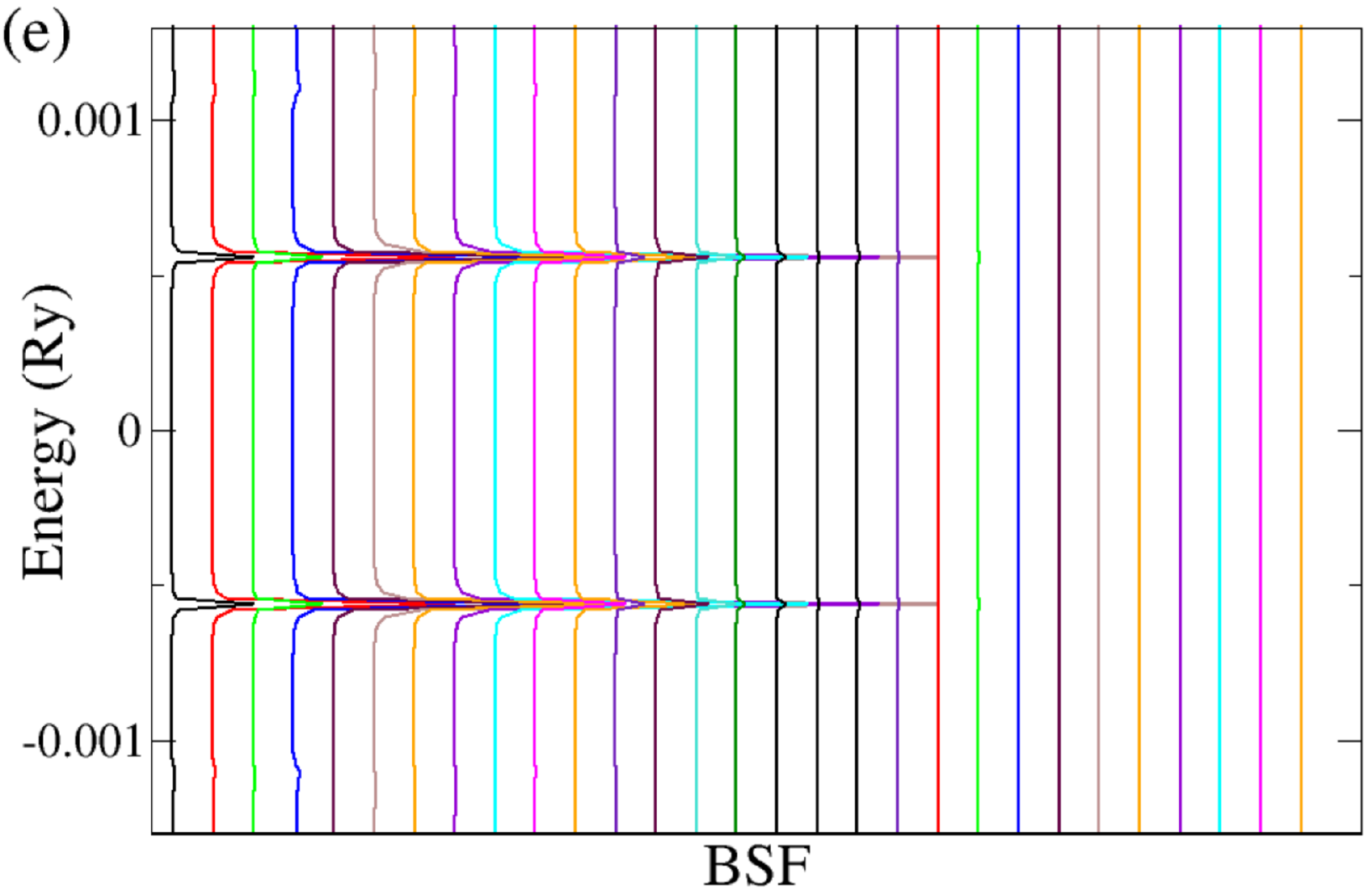}

\vspace{0.2truecm}

\includegraphics[width=0.58\textwidth]{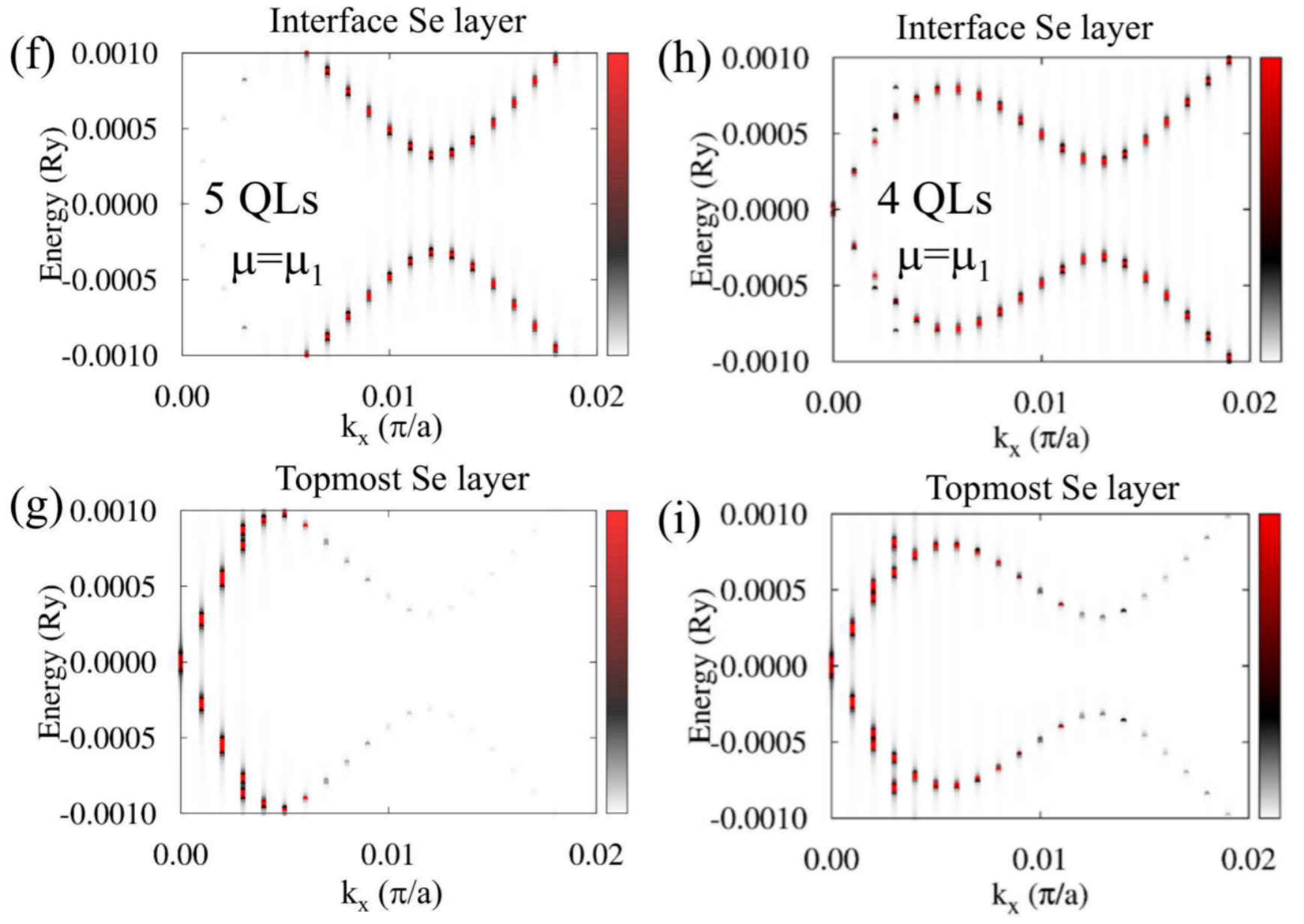}
\hspace{0.2truecm}
\includegraphics[width=0.27\textwidth]{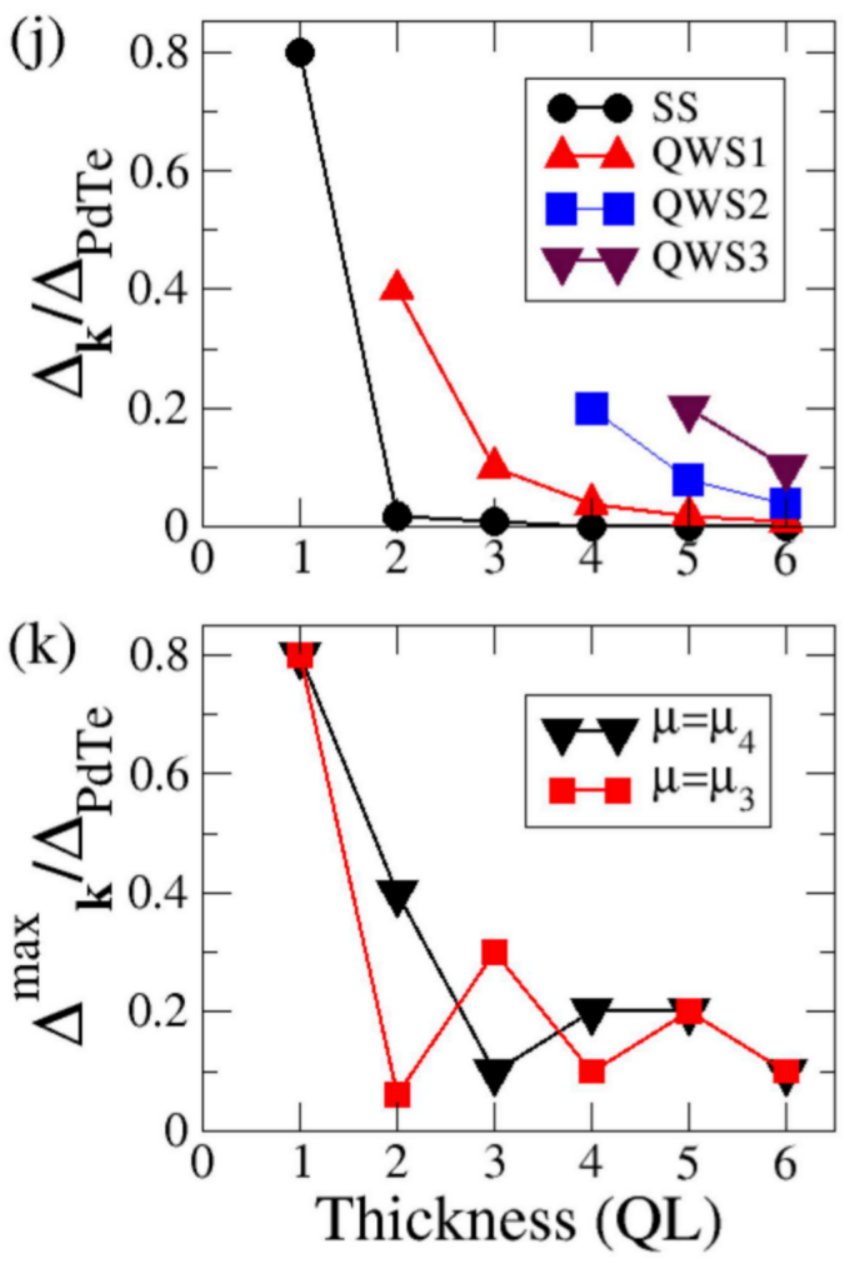}
\caption[Delta]{(a) Normalized SC-state and normal-state BSF vs $z$ at fixed energies and $k_x$ (near $k_1$).
The black (red) curve here corresponds to the black (red) BSF peak at $-0.00032$ ($-0.00052$)~Ry in Fig.~\ref{fig:SC_BSF}(a).
(b) Normalized SC-state BSF vs $z$ at fixed $k_x$ and energies with $\mu_4$ near
$\Delta_{\bf k3}$, $\Delta_{\bf k4}$, $\Delta_{\bf k5}$, $\Delta_{\bf k6}$, and $\Delta_{\bf k7}$.
(c)-(e) Layer-resolved SC-state BSF vs energy at $k_1$ with $\mu_4$ (Table~1), and at $k_x=0.01369$~and 0.03700~$\pi/a$ with $\mu_2$ for
5-QL/PdTe, respectively. The BSF of each layer is shifted and scaled. (f)-(i) BSF contours of the 5-QL and 4-QL overlayers with $\mu_1$,
respectively. (j) Induced gap from different states vs thickness at $\mu_4$. (k) $\Delta^{\rm{max}}_{\bf k}/\Delta_{\rm{PdTe}}$ vs thickness
at $\mu_3$ and $\mu_4$. See the main text for the definition of $\Delta^{\rm{max}}_{\bf k}$.}
\label{fig:gap}
\end{center}
\end{figure*}

Let us now consider that PdTe is in the SC state with a bulk SC gap $\Delta_{\rm{PdTe}}$ of 0.001 Ry and that the TI pairing potential 
$\Delta^{\rm{TI}}_{\rm{eff}}$ is zero.
We then calculate BSF of the heterostructures at a fixed chemical potential. See the SM for the detailed procedure. Figure~\ref{fig:SC_BSF} shows BSF contours of the 5-QL TI overlayer on {\it SC} PdTe as a function of energy and $k_x$ at chemical potential $\mu_4=-0.254$~Ry.
%Our PdTe gap size is about one order of magnitude greater than the experimental gap \cite{Karki2012}, yet it is small enough
%such that our results can be applied to experimental systems.
Within the bulk SC gap, we find a highly momentum-dependent proximity-induced gap feature, $\Delta_{\bf k}$. In particular, we focus on {\it seven} induced gap sizes listed in Table~\ref{tab:mu4}. Two large gaps
$\Delta_{\bf k1}$ and $\Delta_{\bf k2}$ are clearly seen in Fig.~\ref{fig:SC_BSF}(a)-(c).
Gap sizes $\Delta_{\bf k3}$, $\Delta_{\bf k4}$, $\Delta_{\bf k5}$, $\Delta_{\bf k6}$, and $\Delta_{\bf k7}$ are zoomed in in Fig.~\ref{fig:SC_BSF}(d)-(f). Except for $\Delta_{\bf k7}$, all six gap sizes are observed in the BSF of all TI layers. For example,
see Fig.~\ref{fig:gap}(c) in the case of $\Delta_{\bf k1}$. Thus, these gap sizes do {\it not} depend on $z$ for a fixed TI-overlayer thickness.
The induced gap size is larger with smaller $k_x$. The dispersion near the
six gap sizes arises from both electron and hole contributions. See Figs.~\ref{fig:SC_BSF}(g),(h),(j),(k),~S3, and~S4.
In the case of $\Delta_{\bf k7}$, however, the dispersion appears only for the topmost QL and it exhibits a positive (negative)
slope only from the electron (hole) contribution. See Fig.~\ref{fig:SC_BSF}(i) and (l).

% TABLE
\begin{table*}[!h]
\centering
\caption{Induced gaps $\Delta_{\bf ki}$ at $k_i$ and characteristics for the 5-QL overlayer on SC PdTe at chemical potential
$\mu_4=-0.2540$ Ry. Due to the numerical accuracy, any gap size $ < 10^{-6}$ Ry is set to zero.
$\Delta_{\rm{PdTe}}$$=$0.001 Ry. The TI pairing potential is set to zero. See Fig.~\ref{fig:SC_BSF}.
SS denotes the top-surface TI Dirac state.}
\begin{tabular}{c|c|c|c|c|c|c|c} \hline \hline
                 &   $i=1$    & $i=2$       & $i=3$        & $i=4$    & $i=5$    & $i=6$    & $i=7$  \\ \hline
$\Delta_{\bf ki}$ (Ry) & $2.0\times10^{-4}$ & $2.8\times10^{-4}$ & $2.2\times10^{-4}$ & $8.0\times10^{-5}$
& $4.0\times10^{-5}$ & $2.0\times10^{-5}$ & 0 \\
$k_i$ ($\pi/a$)        &  0.05550         & 0.05700           & 0.06930            &  0.06960       &  0.07370       &  0.07380       & 0.08023 \\
origin           & QWS3 & QWS3 & QWS2 & QWS2 & QWS1 & QWS1 & SS \\ \hline \hline
\end{tabular}
\label{tab:mu4}
\end{table*}

The dispersion near the first six induced gaps suggests that Cooper pairs of PdTe tunnel into the TI region, giving rise to the proximity-induced gaps and Andreev bound states. The momentum-dependence and gap sizes can be strongly modified with the interface type and the band structure of non-SC. By comparing the SC-state BSF to the normal-state BSF, we identify that
$\Delta_{\bf k1}$ and $\Delta_{\bf k2}$ originate from Cooper-pair tunneling into QWS3 and that $\Delta_{\bf k3}$ and $\Delta_{\bf k4}$ from
QWS2. Similarly, $\Delta_{\bf k5}$ and $\Delta_{\bf k6}$ arise from Cooper-pair tunneling into QWS1, and $\Delta_{\bf k7}$ from the top-surface
Dirac state. See also Fig.~S5. To elucidate the origin of the two different gap sizes from each QWS, we plot normalized SC-state and normal-state BSF at fixed energies and $k_x$ points near the gap size $\Delta_{\bf k1}$. As shown in Fig.~\ref{fig:gap}(a), the two SC-state BSF qualitatively differ from each other, while the two normal-state BSF are indistinguishable.
The state with a larger spectral weight near the interface gives rise to a larger induced gap. Figure~\ref{fig:gap}(c) also clearly shows the two induced gaps at $k_1$. This two-gap feature is consistent with that of the normal-state BSF (Fig.~S6).
The induced gap size overall increases as Cooper pairs tunnel into higher-energy QWS because of stronger coupling with the substrate. See Fig.~\ref{fig:gap}(b) and Table~1.
The dispersion near $\Delta_{\bf k7}$ (Fig.~\ref{fig:SC_BSF}(i) and (l)) is quite distinct from that near the other gaps.
%This dispersion appears only for the topmost and electron-only or hole-only dispersion is found near the gap.
Cooper pairs do not seem to efficiently tunnel into the top-surface Dirac state which is strongly localized at the top surface, giving rise to
zero induced gap.

We now investigate the effect of chemical potential on the proximity-induced gap at the 5-QL overlayer by considering $\mu_3$, $\mu_2$, and $\mu_1$,
as indicated in Fig.~\ref{fig:Normal_BSF}(g). For $\mu_3$ five distinct gap sizes are found. Similarly to the case of $\mu_4$, the smallest
gap is zero from the top-surface Dirac state, while the rest four gaps are as significant as those for $\mu_4$. The dispersion near the four gaps
has characteristics of Andreev states. As chemical potential decreases, the number of gaps decreases, whereas the gap size associated with
a given QWS type (such as QWS1, QWS2) increases. See Fig.~S7.
For $\mu_2$ and $\mu_1$, only two gap sizes are observed, and Andreev states exist
only near the larger gap. Figure~\ref{fig:gap}(d) and (e) show layer-resolved BSF with $\mu_2$, and Fig.~\ref{fig:gap}(f) and (g) show BSF of the interface and topmost layers with $\mu_1$. The two gaps appear only closer to the interface QL or the topmost QL. The highly suppressed slope of
the dispersion does not allow Cooper pairs to tunnel into the interface TI state until chemical potential reaches near the Dirac points.
For $\mu_2$ and $\mu_1$, only top-surface and interface TI states are involved. The induced gap from
the interface TI state is largest among the gaps found.

% thickness dependence here

To examine the effect of TI film thickness, we plot the smaller induced gap from each TI-state type (i.e., QWS1, QWS2, QWS3, interface or top-surface) vs thickness at $\mu_4$ [Fig.~\ref{fig:gap}(j)]. We find that the induced gap from higher-energy QWS
decays much slowly than that from lower-energy QWS, and that the gap from the top-surface Dirac state becomes zero for thickness greater than 3 QLs.
Figure~\ref{fig:gap}(k) shows the "maximum" gap from the gap sizes shown in Fig.~\ref{fig:gap}(j) as a function of overlayer thickness, at $\mu_3$
and $\mu_4$. Interestingly, the "maximum" gap oscillates with thickness. For overlayers thinner than 5 QLs, the
surface-hybridization effect is also seen in the induced gap. The gap size from the top-surface Dirac state becomes noticeable, and the
induced gap sizes do not depend on $z$. Compare Fig.~\ref{fig:gap}(h) and (i) with Fig.~\ref{fig:gap}(f) and (g) for $\mu_1$.
Tables in the SM list all induced gap sizes.

% Now with effective pairing potential for the TI region not zero.

\begin{figure*}[!h]
\begin{center}
\includegraphics[width=0.7\textwidth]{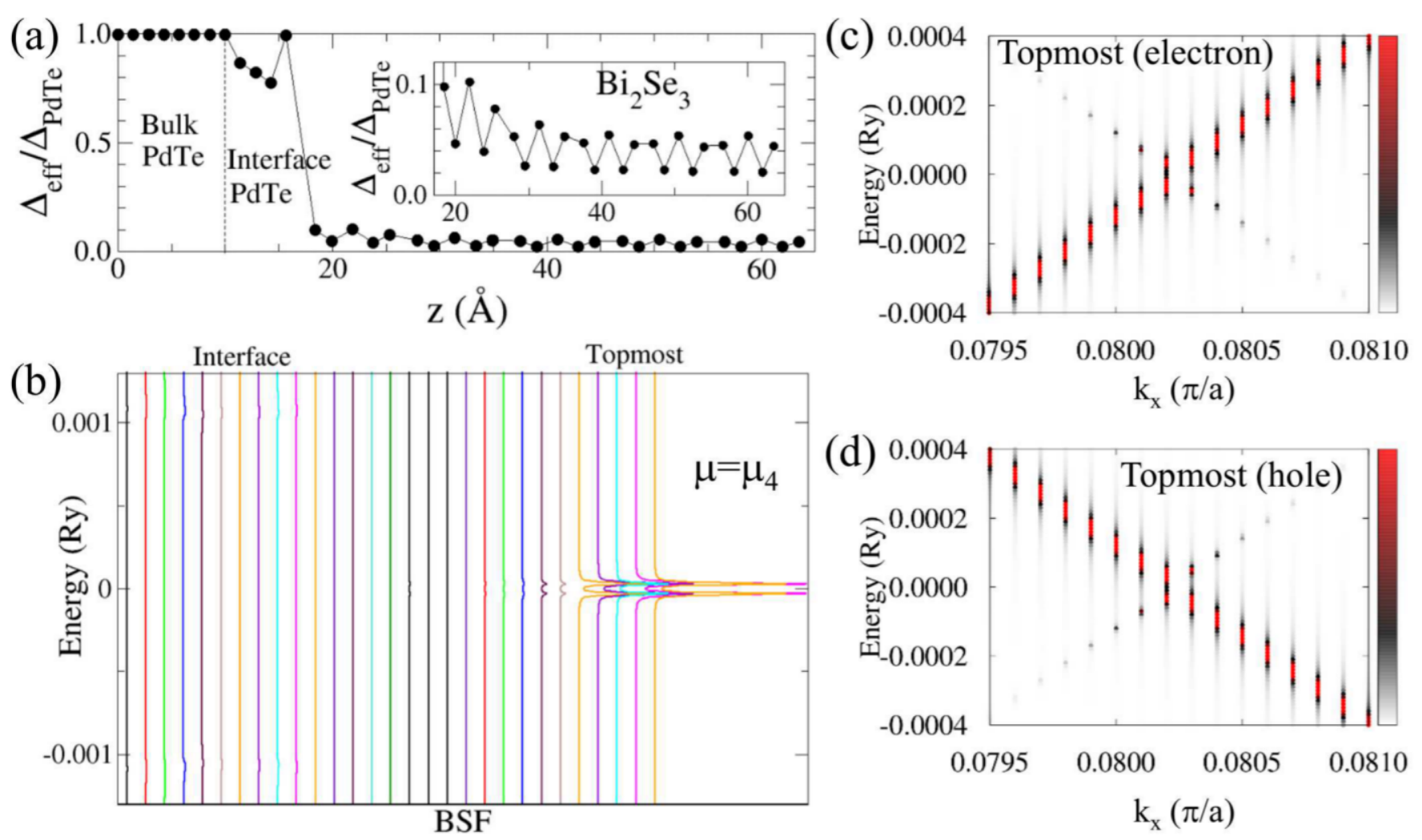}
\caption[SCFgap]{(a) Self-consistently calculated effective pairing potential vs $z$ for the 5-QL/PdTe with the TI region zoomed-in.
(b) Layer-resolved BSF for the 5-QL/PdTe at $k_x=k_4$ with $\mu_4$. (c),(d) Electron and hole BSF of the topmost
Se layer corresponding to (b).
In (b)-(d) the pairing potential in (a) is used.}
\label{fig:SCFgap}
\end{center}
\end{figure*}

\subsubsection{Induced spectral gap with $\Delta^{\rm{TI}}_{\rm{eff}}\neq 0$}\label{sec3:gap2}

We study an effect of the TI pairing potential $\Delta^{\rm{TI}}_{\rm{eff}}$ by keeping $\Delta_{\rm{PdTe}}$ same as before and considering that $\Delta^{\rm{TI}}_{\rm{eff}}\neq 0$. Although electron-phonon coupling of the TI cannot induce superconductivity itself, it can create a finite
pairing potential induced by the superconductor \cite{Fu2008}. Figure~\ref{fig:SCFgap}(a) shows effective pairing potential $\Delta_{\rm{eff}}$ calculated self-consistently for the 5-QL overlayer on PdTe using semi-phenomenological parameters within the SKKR method. See the SM for the 
detailed procedure. As shown in the inset, the pairing potential decays very slowly in the TI region. With this calculated pairing potential, 
we obtain BSF for the 5-QL overlayer on SC PdTe at $\mu_4$. As listed in Table~\ref{tab:SCFgap}, the induced gap from the top-surface Dirac state
becomes noticeable with Andreev-state characteristics (Fig.~\ref{fig:SCFgap}(b)-(d)), whereas the gap from high-energy QWS does not change much
compared to the case of zero TI pairing potential. We expect that topological edge states within the smallest induced gap can trap Majorana
zero-modes with broken time-reversal symmetry \cite{Fu2008,Alicea2012} and that pairing type of the proximity-induced superconductivity may be identified from gap anisotropy \cite{Alicea2012} and/or unique spin-orbital textures of the Andreev states \cite{Fukaya2019}.

% TABLE
\begin{table}[h!]
\centering
\caption{Induced spectral gaps (in units of the bulk SC gap, $\Delta_{\rm{PdTe}}=0.001$~Ry) for the 5-QL overlayer on SC PdTe at chemical
potential $\mu_4=-0.2540$ Ry with two different TI pairing potentials $\Delta^{\rm{TI}}_{\rm{eff}}$.
A smaller gap is chosen when two gaps are found at $k_x$ values close to each other.
QWS1, QWS2, QWS3, and SS denote the first lowest, second, and third QWS in the TI conduction band region
and the top-surface TI Dirac state, respectively. Due to the numerical accuracy, any gap size $< 10^{-6}$ Ry is set to zero.}
\begin{tabular}{c|c|c}
\hline \hline
       &   $\Delta^{\rm{TI}}_{\rm{eff}}\neq 0$ [Fig.~\ref{fig:SCFgap}(a)] &  $\Delta^{\rm{TI}}_{\rm{eff}}=0$ \\ \hline
QWS3   & 0.220        & 0.200  \\
QWS2   & 0.110        & 0.080  \\
QWS1   & 0.056        & 0.016  \\
SS     & 0.030        & 0              \\ \hline \hline
\end{tabular}
\label{tab:SCFgap}
\end{table}

\subsection{Comparison with experiment}\label{sec3:exp}
% Comparison with experiment and other theoretical studies.

Let us make brief qualitative comparison of our results with experimental and model-Hamiltonian studies, considering that our SC substrate
differs from those used in experiments. Refs.~\onlinecite{SYXu2014,Flototto2018} showed momentum-resolved proximity-induced gaps using ARPES.
In Ref.~\onlinecite{SYXu2014} the proximity-induced gap from the bulk TI state significantly differs from the gap from the TI surface state,
while in Ref.~\onlinecite{Flototto2018} that is not the case. Our results are consistent with the former experimental data whether $\Delta^{\rm{TI}}_{\rm{eff}}$ is set to zero or not.
The discrepancy may originate from the experimental difficulty in identifying the peak center of the spectral function corresponding to the
bulk and surface states. We find that an accurate estimate of the induced gap from the TI surface state requires a high precision in momentum
such as 10$^{-5}$~$\pi/a$. With a precision of~0.001 $\pi/a$, the induced gap can be overestimated by two orders of magnitude. However,
the induced gap from bulk states is insensitive to the precision of momentum. Compared to STM experimental data \cite{MXWang2012,JPXu2014}, we obtain underestimated induced gaps even for $\Delta^{\rm{TI}}_{\rm{eff}}\neq 0$. Model-Hamiltonian studies
predicted the zero gap from the top-surface TI state for thick TI overlayers \cite{Stanescu2010,CKChiu2016}, which is consistent with our
result. However, the rich features of the induced gaps that we find were not obtained from the model-Hamiltonian approaches.

%%%%%%%%%%%%%%%%%%%%%%%%%%%%%%%%%%%%%%%%%%%%%%%%%%%%%%%%%%%%%%%%%%%%%%%%%%%
\section{Conclusion}

In summary, we simulated Bi$_2$Se$_3$ films of 1-6 nm overlaid on SC PdTe by solving the DBdG equations with two pairing potential profiles
within the SKKR method, finding that the strong $k_{\parallel}$-dependence of the proximity-induced gap arises from the unique TI band structure
and its modifications under the SC substrate. The size of the induced gap strongly varies with characteristics of TI states which are
partially occupied at the Fermi level. Cooper pairs tunnel into higher-energy QWS more efficiently and the induced gap from higher-energy QWS
decreases more slowly with increasing TI film thickness. For thick TI films, the induced gap from the top-surface Dirac state becomes zero with
zero TI pairing potential, whereas the gap can be substantial for finite TI pairing potential. For a given thick TI film, the induced gap from the interface Dirac state appears only near the interface QL, although the induced gap size from each QWS does not depend on $z$. Our findings
demonstrate an importance of consideration of the realistic TI band structure for studies of the superconducting proximity effect, and they
can be used for future larger-scale simulations and experimental studies of topological edge states or nanowires on SC substrates in pursuit
of Majorana zero-modes.

%%%%%%%%%%%%%%%%%%%%%%%%%%%%%%%%%%%%%%%%%%%%%%%%%%%%%%%%%%%%%%%%%%%%%%%%%%%%%%

%\added{{\noindent {\fontsize{16}{20}{\bf Note}}}}
%\added{{\noindent The authors declare no competing financial interests.}}

%acknowledgments in revtex
%acknowledgment in achemso
\begin{acknowledgements}
We are grateful to Peter Rakyta for setting up initial Bi$_2$Se$_3$ structures. The computational support was provided by San Diego Supercomputer
Center under DMR060009N and Virginia Tech Advanced Research Computing.
BU was supported by the Hungarian National Research, Development and Innovation Office under Contract No. K131938 and
BME Nanotechnology FIKP grants. GC acknowledges support from the European Unions Horizon 2020 research and innovation programme under the Marie Sklodowska-Curie grant agreement No. 754510.
\end{acknowledgements}

\clearpage

%\bibliography{references_v03}


\end{document}